\documentclass[journal,twoside]{IEEEtran}

\usepackage{xcolor}
\definecolor{accessblue}{cmyk}{1,.32,0,.25}
\definecolor{lightskyblue}{HTML}{66C7F4} 
\definecolor{darkblue}{HTML}{1B4079} 
\definecolor{greycolor}{cmyk}{0,0,0,.8}
\definecolor{grey}{cmyk}{0,0,0,.1}
\definecolor{black}{cmyk}{0,0,0,1}
\definecolor{chart1}{HTML}{CA472F} 
\definecolor{chart2}{HTML}{9DD866} 
\definecolor{chart3}{HTML}{6F4E7C} 
\definecolor{chart4}{HTML}{F6C85F} 
\definecolor{chart5}{HTML}{EF709D} 

\usepackage{balance}
\usepackage[caption=false,font=footnotesize]{subfig}
\usepackage{amsmath}
\usepackage{cleveref}
\usepackage{siunitx}
\DeclareSIUnit\dBi{dBi}

\usepackage{url}
\usepackage{algorithm}
\usepackage{algpseudocode}
\usepackage{pgfplots}
\usepackage{pgfplotstable}
\pgfplotsset{compat=1.7}
\usepackage{tikz}
\usetikzlibrary{external}
\let\svtikzpicture\tikzpicture
\def\tikzpicture{\noindent\svtikzpicture}

\usepackage{booktabs}
\usepackage{acronym}
\usepackage{graphicx}
\graphicspath{{./img/}}

\usepackage{array}
\newcolumntype{P}[1]{>{\raggedright\arraybackslash}p{#1}}
\newcolumntype{L}[1]{>{\raggedleft\arraybackslash}p{#1}}

\acrodef{RIS}{Reconfigurable Intelligent Surface}
\acrodef{VNA}{Vector Network Analyzer}
\acrodef{RF}{Radio Frequency}
\acrodef{VNA}{Vector Network Analyser}
\acrodef{HAR}{Human Activity Recognition}
\acrodef{CNN}{Convolutional Neural Network}
\acrodef{HAR}{Human Activity Recognition}
\acrodef{HPR}{Human Posture Recognition}
\acrodef{HGR}{Hand Gesture Recognition}
\acrodef{SoI}{Space-of-Interest}
\acrodef{CS}{Compressive Sensing}
\acrodef{LoS}{Line-of-Sight}
\acrodef{NLoS}{Non-Line-of-Sight}
\acrodef{RSSI}{Received Signal Strength Indicator}
\acrodef{CSI}{Channel State Information}
\acrodef{RFID}{Radio Frequency Identification}
\acrodef{LNA}{Low Noise Amplifier}
\acrodef{FCAO}{Frame Configuration Alternating Optimization}
\acrodef{ANN}{Artificial Neural Networks}
\acrodef{VISA}{Virtual Instrument Software Architecture}
\acrodef{SNR}{Signal-to-Noise Ratio}
\acrodef{FBP}{Feed Beam Point}

\newcommand{\minus}{\scalebox{0.75}[1.0]{$-$}}

\begin{document}
\title{Human Activity Recognition with a \SI{6.5}{\giga\hertz} Reconfigurable Intelligent Surface for Wi-Fi 6E}

\author{Nuno Paulino,~\IEEEmembership{Member,~IEEE,}
        Mariana~Oliveira,~
        Francisco~Ribeiro,~
        Luís~Outeiro,~
        Pedro~A.~Lopes,~
        Francisco~Vilarinho,~
        Sofia~Inácio,~
        Luís~M.~Pessoa,~\IEEEmembership{Member~IEEE}%
\thanks{All authors are with INESC TEC (Institute for Systems and Computer Engineering, Technology and Science) and with the Faculty of Engineering, University of Porto, (Portugal). e-mail: \{nuno.m.paulino, mariana.s.fonseca, francisco.m.ribeiro, luis.outeiro, pedro.a.lopes, francisco.g.vilarinho, sofia.i.inacio, luis.m.pessoa\}@inesctec.pt}.}

\markboth{18.º Congresso do Comité Português da URSI}%
{Paulino \MakeLowercase{\emph{et al.}}, Human Activity Recognition with a \SI{6.5}{\giga\hertz} Reconfigurable Intelligent Surface for Wi-Fi 6E}

\maketitle

\begin{abstract}
\ac{HAR} is the identification and classification of static and dynamic human activities, which find applicability in domains like healthcare, entertainment, security, and cyber-physical systems. Traditional \ac{HAR} approaches rely on wearable sensors, vision-based systems, or ambient sensing, each with inherent limitations such as privacy concerns or restricted sensing conditions. Recently, \ac{RF}-based \ac{HAR} has emerged, relying on the interaction of \ac{RF} signals with people to infer activities. \acp{RIS} offers significant potential in this domain by enabling dynamic control over the wireless environment, thus enhancing the information extracted from \ac{RF} signals.

We present an \ac{HGR} approach that employs our own \SI{6.5}{\giga\hertz} \ac{RIS} design to manipulate the \ac{RF} medium in an area of interest. We validate the capability of our \ac{RIS} to control the medium by characterizing its steering response, and further we gather and publish a dataset for \ac{HGR} classification for three different hand gestures. By employing two \acp{CNN} models trained on data gathered under random and optimized \ac{RIS} configuration sequences, we achieved classification accuracies exceeding \SI{90}{\percent}.
\end{abstract}


\section{Introduction}

\begin{figure}[t!]
    \centering
    \includegraphics[width=1.0\linewidth]{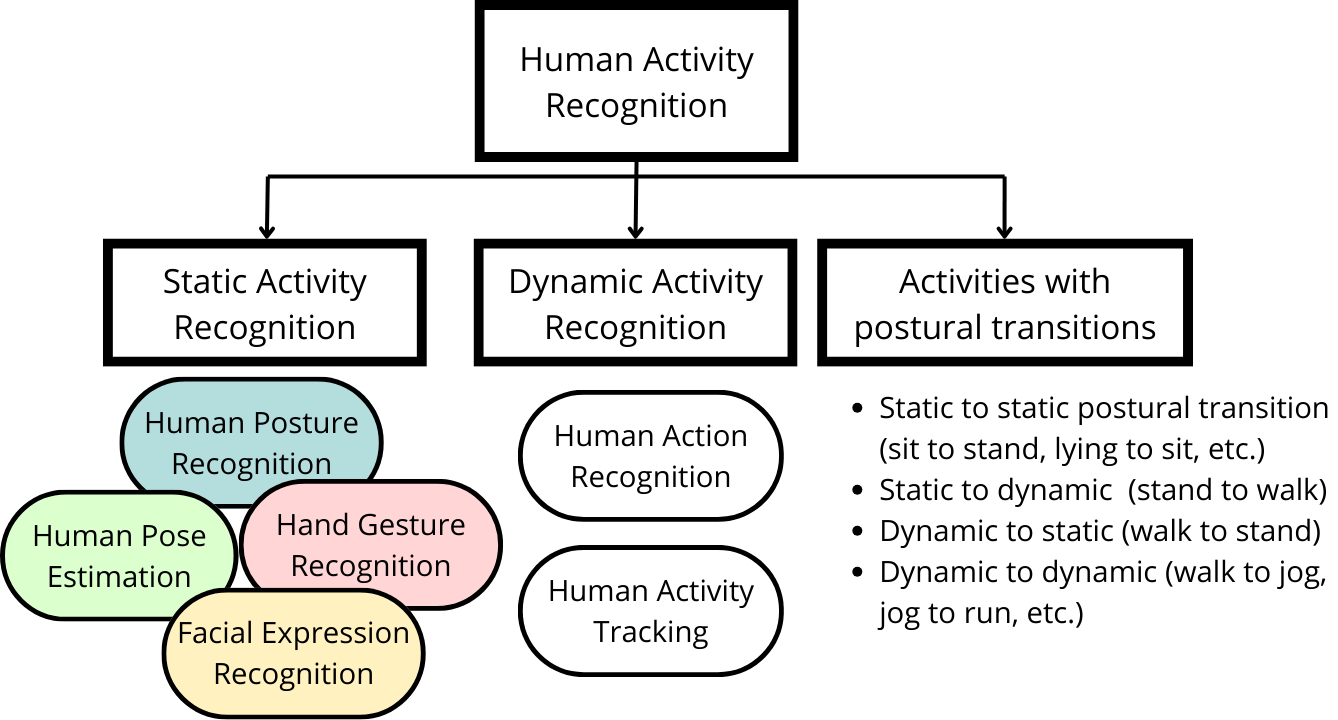}
    \caption{\label{fig:humanactivitytypes} Types of Human Activity Recognition and respective examples.}
\end{figure}

\ac{HAR} is a growing field dedicated to identifying and classifying all forms of human activity, encompassing both static and dynamic motion \cite{s22176463}. \Cref{fig:humanactivitytypes} illustrates different types of human activity, which include static activities, dynamic activities, and activities involving postural transitions \cite{Ahad2021}. Static activities include lying, sitting, and standing, which are generally easier to identify but can be confused with similar postures. Dynamic activities involve continuous movement, such as walking or running, which require distinguishing between different motion patterns. Activities with postural transitions, such as sitting to standing or walking to jogging, involve shifts between static and dynamic states and can be particularly complex due to overlapping motion patterns.

The capability of performing accurate \ac{HAR} has the potential to significantly impact various domains \cite{7975660}. In healthcare, \ac{HAR} can assist in monitoring patient conditions and detecting early signs of medical issues \cite{10.1007/978-3-031-24352-3_5}. In smart environments, \ac{HAR} can enhance home assistance services to improve quality of life and autonomy \cite{s21186037}. In sports, it can contribute to improving personal health and well-being \cite{s19225001}. In security, \ac{HAR} can be used for surveillance to identify unusual events \cite{9616791}. Additionally, \ac{HAR} improves human-computer interaction (HCI) by enabling more intuitive and responsive interfaces \cite{xu2017realtimehandgesturerecognition}.

However, \ac{HAR} is complex, requiring selecting and deploying suitable sensors to collect information-rich data, pre-processing the data, and extracting relevant features. Afterward, a range of approaches can be considered to process the data, including machine learning and deep learning algorithms to recognize activities \cite{hussain2020review}. Regarding specific sensor types, \ac{HAR} approaches may be vision-based, through monocular or stereo cameras \cite{JALAL2017295}, based on sensors such as accelerometers, which can be worn on the body or integrated into smartphones or smartwatches to capture direct movement data \cite{9176239}. Additional sensors can provide secondary parameters like temperature and humidity, and provide further contextual information \cite{roy2016ambient}. 

Another emergent method for \ac{HAR} is the use of \ac{RF} characteristics of the environment, where data relative to the interaction of stray or controllable \ac{RF} signals with objects or people within a given \ac{SoI} is used as the sensor component of the system. By relying on \ac{RF} signals, activities can be detected or classified through the effects of body movements on the propagation of these signals, allowing for \ac{HAR} techniques that address concerns about monitoring and privacy \cite{YANG2024100204}. It does not require wearable or environmental sensors, providing a noninvasive and lower cost approach, instead exploiting the already existing plethora of \ac{RF} signals in the environment that originate from ubiquitous commodity networks like 5G or WiFi. Since reflected \ac{RF} signals are also captured by receiving antennas, information can be gathered even in \ac{NLoS} conditions \cite{9982449}, and depending on the signal frequency, even through-wall detection can be considered \cite{8578866,10.1145/2816795.2818072}. Finally, the use of higher frequency signals can even provide a high spatial resolution, which has demonstrated advantages over vision-based systems for multi-person recognition \cite{10155196}.

Despite these advantages, wireless environments are complex and difficult to model. Not only is there no direct control over the stray \ac{RF} signals in a region of interest, but there is also the complexity of multi-path interference, caused when signals reflect off different surfaces and reach the receiver through various paths \cite{wang2024generativeairfsensing}. Thus the difficulty of clearly relating and modeling the parameters of received \ac{RF} signals to the status of the environment derives from a largely uncontrolled propagation medium. 

\begin{figure}[!t]
    \centering
    \subfloat[Regular Refletor]{\includegraphics[width=0.50\linewidth]{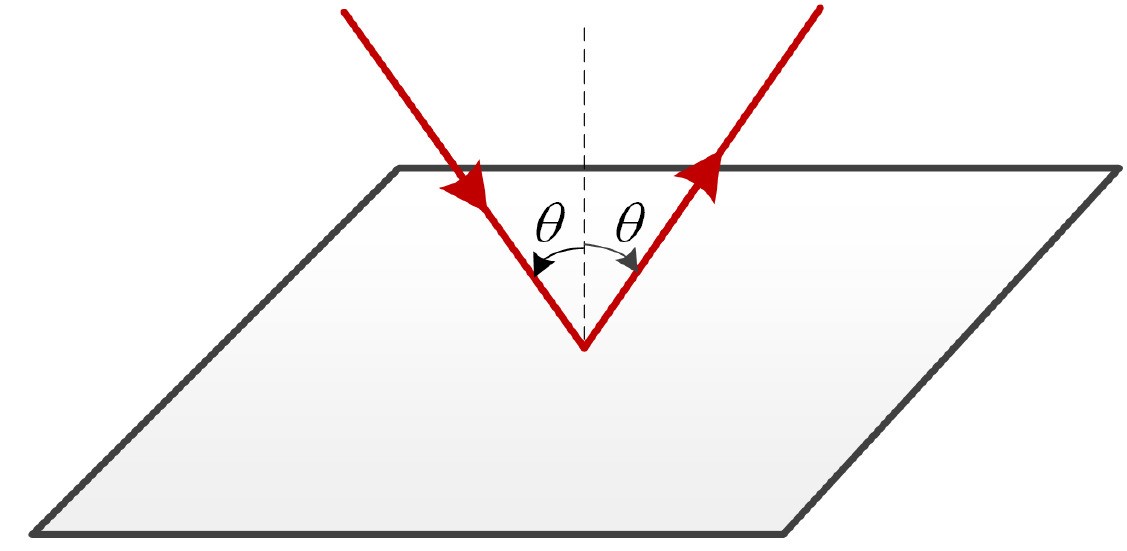}\label{spec_dif}}
    \subfloat[\ac{RIS}]{\includegraphics[width=0.50\linewidth]{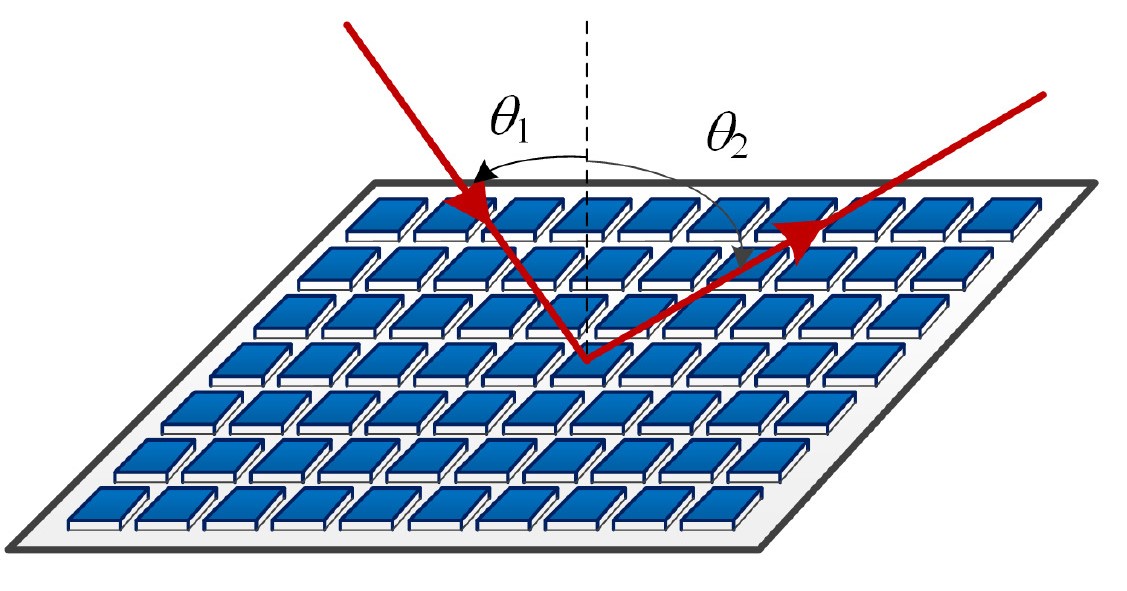}\label{ris_dif}}
    \caption{Regular reflector vs. \ac{RIS}, which achieves non-specular reflections through programmable phase shifts (adapted from \cite{image_ris_metal}).}
    \label{angles_Refle}
\end{figure}

\acfp{RIS} are a promising solution to achieve this medium control. They are composed by a 2D matrix of antenna elements, also called unit cells. These cells contain no active \ac{RF} chains and act only as reflectors. Simple electronics are used to change the reflection coefficient of each individual unit cell, and consequently a \ac{RIS} can adjust the surface impedance to produce specific phase shifts for each of the reflected replicas of the incident signal, thus achieving control of the wireless environment \cite{9424177,Zhang2021}. 
    
For instance, whereas a specular reflector like a metal surface shown in \Cref{spec_dif} deflects a signal at an angle equal to the angle of incidence, specific control patterns applied to the unit cells of a \ac{RIS}  can be used to generate constructive interference to focus the total reflection into a particular direction, achieving non-specular reflections, as shown in \Cref{ris_dif}.

The advantages for communication systems are increased spectral efficiency and power consumption, by reducing radiation towards unintended directions, and support for \ac{NLoS} communications, since links can be redirected around obstacles. But applications go beyond communication, as this medium control places \acp{RIS} as a candidate technology to also enable applications like localization, imaging, or sensing \cite{10.1145/2785956.2787487,9508872,9874802,9424177,10.1109/MWC.011.2100016,9133157}, in emergent networks like 6G \cite{10352433}.

Concurrently, the WiFi-6E standard has recently been defined, promising to deliver a throughput of up to \SI{9.6}{Gbps} by extending WiFi capability into the \SI{5.925}{\giga\hertz} to \SI{7.125}{\giga\hertz} range. This expansion allows for up to seven \SI{160}{\mega\hertz} MHz channels, compared to the two channels available in Wi-Fi 5 and Wi-Fi 6, or alternatively up to 59 channels with a bandwidth of \SI{20}{\mega\hertz}. Despite this, this frequency range suffers from greater path loss relative to previous WiFi standards, and also greater attenuation through walls or objects, compromising \ac{NLoS} operations. This introduces the need for signal repeaters or greater transmission power.   

Considering the signal manipulation capabilities explained before, namely signal steering, \acp{RIS} are once again placed as a potential low-cost solution to fulfill the expected benefits of WiFi-6E. Moreover, the use of \acp{RIS} operating in WiFi ranges has already demonstrated the viability of localization \cite{10.1145/3625687.3625806} and imaging \cite{9906891} systems. However, to the best of our knowledge, \ac{RIS} designs for Wi-Fi 6E are under-explored. Given the anticipated ubiquity of Wi-Fi networks and the advantages offered by Wi-Fi 6E, we address this gap by investigating 
the use of a Wi-Fi 6E \ac{RIS} for sensing applications.

In this paper, we implement an \ac{HAR} recognition approach by using a \ac{RIS} to implement sensing of the medium, allowing for activity classification through \ac{RF} channel data. Specifically, we will address the classification of static activities, namely \ac{HGR}, which focuses on detecting hand movements and static hand positions. This technique is relevant for applications like human-computer interaction, virtual reality, and sign language recognition \cite{9622242}.

The resulting contributions are as follows: 
\begin{itemize}
 \item The design and fabrication of a PIN diode based unit cell for a \ac{RIS} and with a 1-bit phase control, targeting the range of WiFi6-E (for $f_c = \SI{6.5}{\giga\hertz}$).
 \item Characterization of the magnitude and phase response of the unit cell in a waveguide environment, and of the radiation pattern of an 8x8 \ac{RIS} tile in an anechoic chamber for different frequencies and steering angles. 
 \item Modeled the environment and channel for an \ac{HGR} system, and implemented \ac{RIS} configuration sequences for sensing the environment through $S_{21}$ parameters.
 \item Collected and made publicly available a dataset of $S_{21}$ data designed for \ac{RF}-based classification of three human hand gestures.
 \item Demonstrated the successful classification of gestures using this dataset, for two different \ac{CNN} architectures.
\end{itemize}

The remainder of this paper is organized as follows: in \Cref{sec:related} we review general aspects of \ac{HAR} and \ac{RIS} design and relevant works in both fields; in \Cref{sec:implement} we explain our approach for \ac{HAR} -- specifically \ac{HGR} -- using our own \ac{RIS} design, including an explanation of our channel model, \ac{RIS} configuration scheme, and gesture classification method; in \Cref{sec:experiments} we present the results of a series of experiments, including characterization of the fabricated \ac{RIS}, and different sensing setups and dataset gathering, culminating in a \ac{CNN} based classification of hand gestures through \ac{RF} data; finally, \Cref{sec:conclusion} concludes the paper.

\section{Related Work}
\label{sec:related}

Achieving high performance \ac{RF} sensing is challenging in general, particularly since the areas of interest for sensing, and the \ac{RF} medium itself, are subject to physical obstacles compromising \ac{LoS}, and the effects of multipath interference. Functional radio limitations such as the viable amount of information to be carried by transmission channels also compromise the viability of sensing. 

To tackle this, some approaches rely on deep learning techniques to extract features directly from \ac{RF} data without explicit modeling of complex physical processes, enhancing data processing capabilities \cite{9374635}. However, the major obstacle in \ac{RF} sensing is the uncontrollability of the medium. The use of \acp{RIS} allows for this control to be achieved, 
as the response of the medium 
is a function of known applied control. 

\begin{figure}[!t]
    \centering
    \subfloat[conventional environment]{\includegraphics[width=0.49\linewidth]{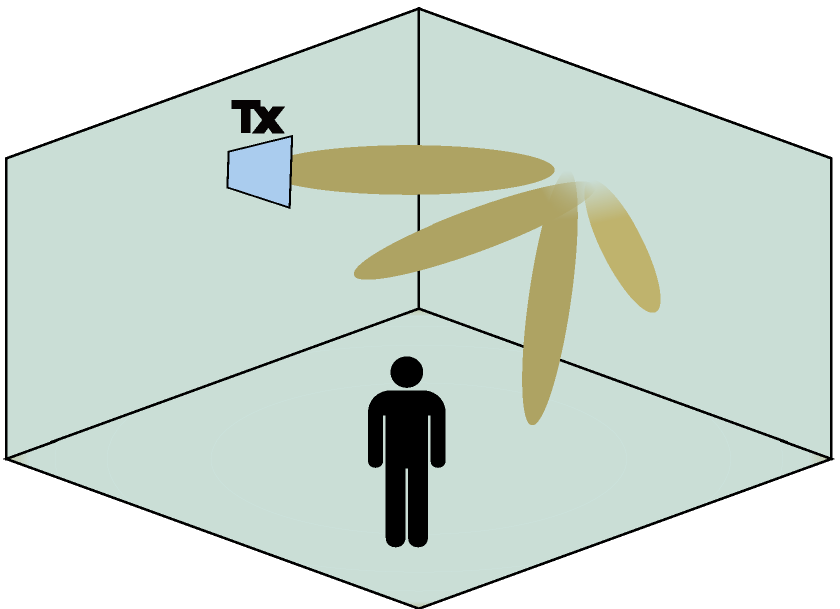}}
    \hfill
    \subfloat[smart environment]{\includegraphics[width=0.49\linewidth]{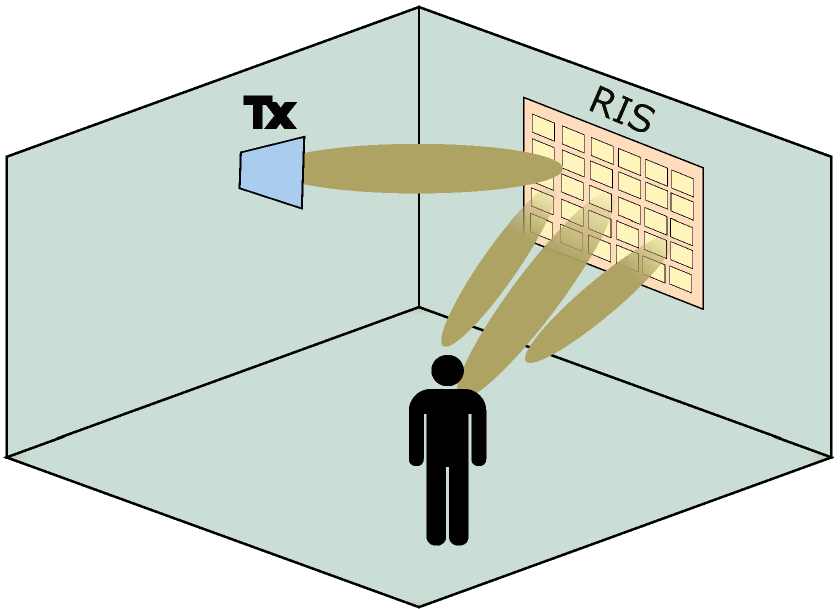}}
    \caption{a) Conventional environment with diffuse signal scattering; b) Reconfigurable environment with \ac{RIS} enabling targeted signal beamforming.}
    \label{smart_envir}
\end{figure}

A \ac{RIS} can manipulate the propagation of electromagnetic waves, enabling dynamic adjustments of phase shifts and amplitude ratios along propagation paths. By reconfiguring these parameters, \ac{RIS} creates more favorable channel conditions for \ac{RF} sensing, optimizing signal transmission and reception \cite{10.1145/3379092.3379102}. Thus, unlike conventional surfaces, which scatter radio waves based on their material properties and boundary conditions, a \ac{RIS} can direct reflected signals towards specific locations, enhancing the accuracy of RF sensing systems for human activity detection, as shown in \Cref{smart_envir}.

Given this context, we now review the state-of-the-art from two perspectives. Firstly we present relevant works in the field of \ac{RF} sensing for \acf{HAR}, focusing specifically on \ac{RIS} based approaches. We also provide a respective summary in \Cref{table:har-ris} at the end of the section, including our own approach. Secondly, we briefly review hardware designs for \ac{RIS} designs comparable to the design we have employed in this paper.

\subsection{RF Sensing through \aclp{RIS}}

In \cite{9133157} the authors propose a novel \ac{RF} sensing system that actively customizes the propagation environment by optimizing the configurations of the \ac{RIS}. This approach addresses the limitations of conventional RF sensing techniques, which often struggle with multi-path fading and restricted transmission channels. By leveraging the capabilities of the \ac{RIS}, the system can create multiple independent paths for signal transmission, thereby enriching the information available for accurate posture recognition. The authors employ the \ac{FCAO} algorithm, which minimizes the mutual coherence of the measurement matrix. This reduction in mutual coherence enhances the system's ability to distinguish between different postures and facilitates this classification task. The experimental setup features a horn antenna that emits signals towards the RIS, while an omnidirectional antenna captures the reflections from the human body, utilizing $S_{21}$ parameters as raw data to evaluate the system's performance. The effectiveness of their approach is demonstrated through practical experiments, resulting in a remarkable improvement in recognition accuracy for 4 postures, achieving up to \SI{96.7}{\percent} with optimized configurations, compared to \SI{82.1}{\percent} in random configurations and \SI{73.2}{\percent} in non-configurable environments.

Li \textit{et al.} \cite{li2019intelligent} present a significant advancement in radio-frequency (RF) sensing systems with their intelligent metasurface imager and recognizer, which operates at approximately 2.4 GHz, one of the standard Wi-Fi frequency bands. This system enhances \ac{HAR} by adaptively manipulating ambient RF signals to capture detailed images of human bodies and recognize gestures and vital signs in real-time. The approach relies on an \acp{ANN} to process RF signals, effectively transforming them into silhouette images of individuals by integrating a vision dataset. The authors adapt the configuration of the \ac{RIS} in real-time, according to the processed RF data and the specific areas of interest identified by the \acp{ANN}. This capability allows the system to focus electromagnetic fields on selected body parts while minimizing interference from the surrounding environment and  enables the system to distinguish between various hand signs and monitor vital signs, such as respiration, even when individuals are uncooperative or positioned behind obstacles. The experimental setup includes a large-aperture programmable metasurface, transmitting and receiving antennas, and a \ac{VNA} for data acquisition, along with a network of \acp{ANN}. This configuration enables the system to distinguish between various hand signs and monitor vital signs, such as respiration, even when individuals are uncooperative or positioned behind obstacles.

In \cite{LI2020100006} the authors present an approach of electromagnetic sensing utilizing a programmable metasurface for imaging and body gesture recognition tasks at 2.4 GHz. This innovative system shapes waves illuminating the scene and enhances sensing quality by employing multiple coding patterns of the metasurface, resulting in improved resolution in the image reconstruction of body postures. A key contribution of this work is the integration of two deep \acp{ANN}: the measurement ANN (m-ANN) for adaptive data acquisition and the reconstruction ANN (r-ANN) for instant data processing. This integrated framework allows for the simultaneous optimization of measurement strategies and data processing schemes, significantly enhancing the efficiency and accuracy of human activity recognition tasks. The experimental setup employs a dual-antenna configuration using horn antennas, where one antenna serves as the transmitter and the other as the receiver. These antennas are connected to a \ac{VNA}, which captures the $S_{21}$ parameters as the data. 

Wang \textit{et al.} \cite{wang2022intelligent} introduce an innovative camera system that integrates programmable electromagnetic metasurfaces with deep learning techniques, establishing a Bayesian inference framework to enhance data acquisition and processing. A key contribution of this work is the development of an intelligent EM metasurface camera capable of visualizing human behaviors behind a 60 cm-thick reinforced concrete wall. This capability is achieved through the use of a large-aperture programmable metasurface, which allows for adaptive data acquisition, and artificial neural networks that facilitate instant data processing. The system operates at a frequency of approximately 2.4 GHz, demonstrating robust performance in real-world settings, even in visually obstructed environments. The experimental setup includes a low-cost commercial software-defined radio device (Ettus USRP X310), a transmitting antenna, and a three-antenna receiver, and a single controlling personal computer. The camera employs a compressive microwave measurement strategy, utilizing 18 control patterns per image to optimize the trade-off between imaging quality and efficiency. This innovative approach not only enhances the frame rate but also significantly improves the \ac{SNR} of the acquired data, making it a promising tool for smart community applications and beyond.

\begin{table}[t!]
\centering
\footnotesize
\caption{Summary of RIS-based \acl{HAR} Approaches \label{table:har-ris}}
\begin{tabular}{L{1.25cm}|P{0.94cm}P{0.68cm}P{1.6cm}P{1.97cm}}
\toprule
\textbf{Approach} & \textbf{Anechoic Env.} & \textbf{Freq. (GHz)} & \textbf{Type of HAR} & \textbf{RIS Config.} \\ \midrule
Hu \textit{et al.} \cite{9133157} & Yes & 3.2 & Posture Recognition & \acs{FCAO} algorithm \\ \midrule
Li \textit{et al.} \cite{li2019intelligent} & No (indoor) & 2.4 & Hand Gesture Recognition, Vital Signs & Gerchberg-Saxton algorithm \\ \midrule
Li \textit{et al.} \cite{LI2020100006} & No (indoor) & 2.4 & Posture Recognition & m-ANN \\ \midrule
Wang \textit{et al.} \cite{wang2022intelligent} & No (indoor) & 2.4 & Behavior Monitoring & Random patterns \\ \midrule
Li \textit{et al.} \cite{li2019machine} & Yes & 2.4 & Posture Recognition  & Gerchberg-Saxton algorithm \\ \midrule
This work & Yes & 5.0 to 6.5 & Hand Gesture Recognition & FCAO Algorithm\\ \bottomrule
\end{tabular} 
\end{table}

Li \emph{et al.} \cite{li2019machine} present a digital imaging system based on a reprogrammable metasurface that utilizes machine-learning techniques to optimize real-time image capture. The system employs a two-step training strategy where desirable radiation patterns are first learned using machine learning algorithms, and then the corresponding coding patterns of the metasurface are designed to achieve the desired radiation configuration. This approach enables the generation of optimized measurement modes that facilitate high-quality object recognition and imaging while significantly reducing the number of required measurements. By leveraging a modified Gerchberg-Saxton algorithm \cite{gerchberg1972practical} for discrete-valued optimization, the system effectively tailors the electromagnetic wave manipulation in real-time, enhancing the accuracy and efficiency of the imaging process. A key contribution of this research is the implementation of an image that can be electronically reprogrammed in real-time, allowing for the generation of measurement modes required for machine learning techniques, such as Principal Component Analysis (PCA). The authors demonstrate that by using a prototype of the imager, it is possible to achieve high-accuracy object classification and image reconstruction, even in dynamic scenarios. The experimental results validate the effectiveness of the system, showing that it can operate with a significantly reduced data acquisition time, representing an important advancement in the field of imaging and recognition. The experimental setup consists of a $ \SI{2}{\meter} \times \SI{2}{\meter} \times \SI{2}{\meter}$ anechoic chamber, equipped with a transmitting horn antenna and a waveguide receiver, with a \ac{VNA} for $S_{21}$ data acquisition. 

Finally, the last row of \Cref{table:har-ris} lists the approach presented in this paper. Relative to the state-of-the-art works we have reviewed, it is, to the best of our knowledge, the only \ac{RF} sensing approach based considering the WiFi-6E frequency range, with or without the use of a \ac{RIS}.

\subsection{Unit Cell and \acl{RIS} Designs}

In \cite{9668918} a \ac{RIS} design for \SI{3.5}{\giga\hertz} is evaluated in a real-world environment. The design is fabricated on \SI{1.52}{\milli\meter} thick F4BT450 substrate and contains 2430 unit cells in a near square layout. A varactor is employed per unit cell to achieve a phase control range of close to \SI{360}{\degree}. The experimental setup consisted of an RX in non-line-of-sight from the TX and used an SDR implementation to stream data using QPSK modulation. The \ac{RIS} and RX were placed at three different relative orientations to demonstrate the  \ac{RIS} steering capabilities. 

A \ac{RIS} tuned for \SI{5.8}{\giga\hertz} with 16x10 elements is presented in \cite{9704068}. Using one PIN diode per element a phase difference of \SI{171}{\degree} between states is achieved. The design uses a single layer Rogers RT 6002 substrate, and the diodes are controlled by shift registers and a micro-controller on an external PCB. The \ac{RIS} is evaluated outdoors, with a feed horn placed \SI{5}{\meter} away on the same horizontal plane as the \ac{RIS}. An RX horn placed \SI{10}{\meter} away is moved in a quarter circle range, verifying that the \ac{RIS} improves the SNR at the RX by up to \SI{15}{\decibel}.

In \cite{10389095} the authors present an open-source design for a \ac{RIS} and unit cell tuned for \SI{5.5}{\giga\hertz}, fabricated on FR4 substrate with three layers. Each unit cell is controlled by a low-power RF switch, which sets the unit cell to either a short or open state. An ideal difference of \SI{180}{\degree} is achieved at the central frequency. The \ac{RIS} contains 256 elements in a 16x16 disposition and is controlled by an additional board assembled to its bottom layer which interfaces via USB or Bluetooth.

In \cite{10473212} focus is on datasets of sub-\SI{6}{\giga\hertz} \ac{RIS} characterization, highlighting that the lack of available datasets hinders research on codebook generation or optimization. They also make available two of their own datasets collected with a 10x10 \SI{5.3}{\giga\hertz} \ac{RIS} on an FR4 substrate with \SI{0.5}{\milli\meter} thickness. The unit cells rely on SP8T RF switch which enables configuring the phase response to eight different values. Using a setup similar to ours, with the \ac{RIS} and feed horn mounted on a turntable, the steering performance is characterized for a range of $\pm$\SI{90}{\degree} and sub-arrays of elements of several sizes.

\section{Implementing \acl{RIS} Based \acl{HGR}}
\label{sec:implement}

Recent work has shown the viability of using $S_{21}$ parameters as data for \ac{HAR}-related classification tasks \cite{har_onbody_ssl,har_onbody_cnn,hand_s21}. This includes our own previous work, where we implemented \ac{HPR} using \ac{RF} sensing \cite{10621437}. However, we relied only on specular reflections, i.e., without resorting to a \ac{RIS}. In this paper, we enhance the approach by using our own \ac{RIS}, which we have designed and fabricated for a central frequency of \SI{6.5}{\giga\hertz}, thus keeping in mind its application for WiFi-6E capable scenarios. The \ac{RIS} enables dynamic and configurable signal paths, providing greater flexibility in RF signal manipulation \cite{dynamic_ris_1,dynamic_ris_2,dynamic_ris_3}.

\begin{figure}[t!]
    \centering
    \includegraphics[width=0.90\linewidth]{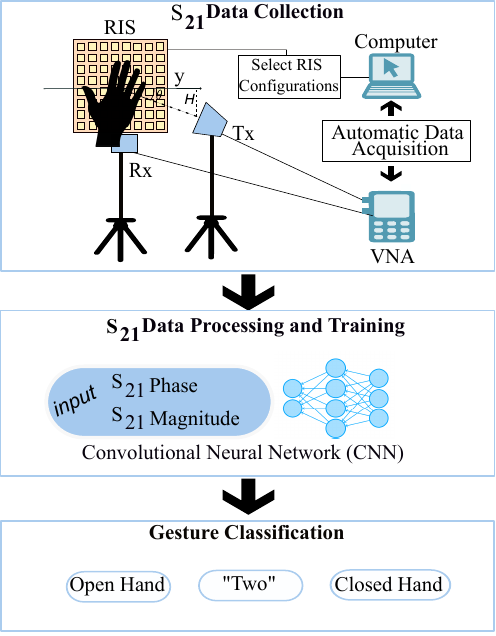}
    \caption{System architecture diagram illustrating the three main components: Data Acquisition, Data Processing and Training, and Gesture Classification. \label{fig:diagram2}}
\end{figure}

Due to the dimensions of our \ac{RIS} tile ($\SI{18.46}{\centi\meter}^2)$, we focus specifically on \acf{HGR}. That is, since the dimension of the \ac{RIS} restricts it to a smaller area of interest, we consider hand gesture recognition a more viable application for this setup.
Our approach and setup are illustrated in \Cref{fig:diagram2}, divided into three stages. Firstly, we design and implement the experimental setup for data collection. This includes the channel between a source antenna Tx, our \ac{RIS} design, and the target receiver antenna Rx. The \ac{SoI}, where the human hand whose gesture we wish to classify is present, is placed in front of the \ac{RIS}. Data is gathered by configuring the \ac{RIS} to a desired state, triggering the transmission of a signal from the Tx through a \ac{VNA}, and recording the $S_{21}$ parameters at the Rx. We gather data for three different hand gestures. To determine the \ac{RIS} configurations to use, we adopted a strategy similar to the one presented by Hu \textit{et al.} \cite{9133157}, as we explain in detail in \Cref{sub:risconfig}.

Secondly, we rely on \acp{CNN} to interpret \ac{RF} signals as images, as presented by Li \textit{et al.} \cite{li2019intelligent}. Thus we process the $S_{21}$ data in order to represent it as image data and train two different \ac{CNN} architectures. Specifically, we train each architecture with two datasets, one gathered using random sequences of \ac{RIS} configurations, and another using optimized sequences which increase the information extracted from the channel. Finally, we evaluate the learning rates and classification accuracies of both models for both datasets.

The following sections present the implementation aspects of each component. 
Firstly, in \Cref{sub:modelsetup}, we explain 
the model of the physical setup of the approach, as well as the channel model including the \ac{RIS}. We then explain in \Cref{sub:risconfig} how sensing can be achieved by then applying a sequence of \ac{RIS} configuration states over this model, and also explain how we generate an optimized sequence of these states. Finally, we explain the hardware implementation aspects, namely of the unit cell and of the \ac{RIS} tile \Cref{sub:risimplement}.

\subsection{Modeling and Configuring the \ac{HGR} Experimental Setup}
\label{sub:modelsetup}

Our approach requires determining a physical configuration between the \ac{RIS}, TX, and RX. Proper antenna positioning is crucial to ensure the quality of the transmitted and received signals, and to maximize the efficiency of signal emission and reception. To achieve this, we require a model of the propagation environment between the \ac{RIS} and \ac{SoI}. We explain both these aspects in the following sections.\\

\begin{figure}[t!]
\centering
\subfloat{
\includegraphics[height=0.315\linewidth]{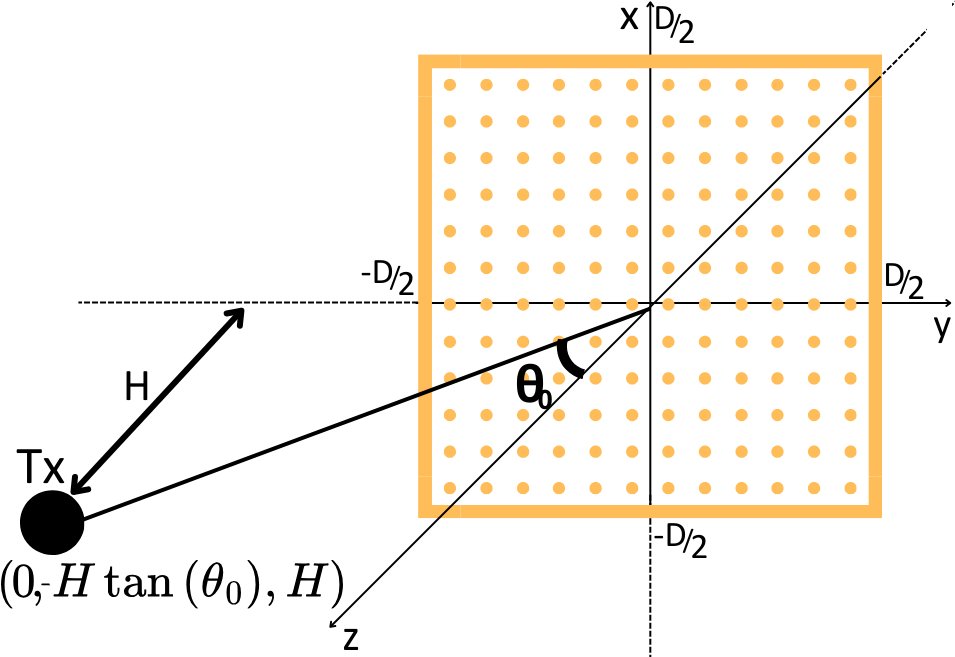}
\label{fig:txsetup1}
}
\subfloat{
\includegraphics[height=0.300\linewidth]{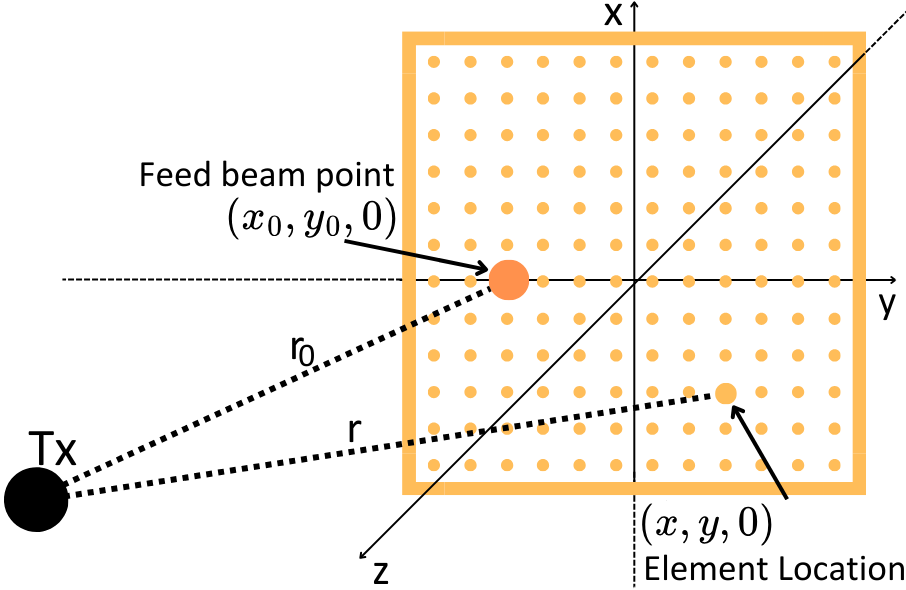}
\label{fig:txsetup2}
}
\caption{Geometric configuration between Tx and \ac{RIS}, considering distance ($H$), offset angle $\theta_0$, and location of \ac{FBP}.}
\label{fig:txsetup}
\end{figure} 

\begin{table}[t!] 
\caption{Parameters for HGR Experimental Setup}
\begin{tabular}{l|l}
\toprule
\textbf{Parameter}                & \textbf{Formula} \\ \midrule
Feed Location                     & $(x_c, y_c, z_c) = (0, -Htan\theta_0, H)$ \\
Feed Beam Point             & $(x_0, y_0, 0)$ \\
Element location                  & $(x, y, 0)$ \\
$d_{f\text{---}FBP}$ & $r_0 = \sqrt{(x_0^2 + y_0^2 + H^2 sec^2(\theta_0) + y_0(2Htan\theta_0)}$\\
$d_{f\text{---}e}$ & $r = \sqrt{(x^2 + y^2 + H^2 sec^2(\theta_0) + y(2Htan\theta_0)}$   \\
$d_{FBP\text{---}e}$& $s = \sqrt{(x - x_0)^2 + (y - y_0)^2}$\\
\bottomrule
\end{tabular} 
\label{tab:table_tx}
\end{table}

\noindent\textbf{Antenna Setup and Physical Layout}
Regarding antenna selection, we considered canonical or pyramidal horns. Pyramidal horn antennas suit our setup best, as they combine E-plane and H-plane sectoral horns for equal radiation patterns \cite{pyramidal}. In contrast, the beam widths in the principal planes E and H of the conical horn are generally unequal \cite[p. 15-2]{book_conical}, which does not meet the symmetrical characteristics required for our setup. For our analysis, we employed a broadband pyramidal horn antenna, specifically an LB-20180-SF, with a frequency range of \SI{2}{\giga\hertz} to \SI{18}{\giga\hertz} and \SI{12}{\dBi} gain. We chose the same horn for the receiver Rx, to ensure that we capture primarily signals from the \ac{SoI}, while minimizing any \ac{LoS} component of the Tx-Rx path.


We then modeled the physical layout between the \ac{RIS} and Tx antenna, as shown in \Cref{fig:txsetup}. The center of the \ac{RIS} is the origin of the coordinate system. The feed antenna Tx is characterized by an offset angle $\theta_0$ measured between itself and the normal vector to the plane of \ac{RIS}, and is situated at a distance $H$ from the xy-plane. Finally, the Tx is directed at a given \ac{FBP}. With this model, we employ the principles regarding aperture efficiency, outlined in \cite{livro_antena_pos}, in order to establish the ideal distances between components. Specifically, we aimed to achieve an edge taper of \SI{-10}{\decibel} to ensure that strong diffractions do not occur at the aperture edge, thus minimizing unwanted radiation and diffraction effects that can degrade the quality of the transmitted signal. To facilitate this analysis and given the configuration shown in \Cref{fig:txsetup}, all the important geometric quantities involved are listed in \Cref{tab:table_tx}, based on the equations outlined in \cite{formulas_ilum}. 

Firstly, we model the radiation of this antenna as a $cos^q$ radiation pattern. To determine the $q$ value, we compared the best fit between datapoints of the radiation pattern of the antenna's datasheet and the $cos^q$ model, at our target central frequency of \SI{6.5}{\giga\hertz}, leading to a choice of $q = 5$.

Secondly, we considered the aperture's spillover efficiency ($\eta_s$) and its illumination efficiency ($\eta_i$). The spillover efficiency quantifies the fraction of power radiated by the feed that is captured by the reflecting aperture, compared to the total radiated power, and is defined as shown in \Cref{eq:eta_s}. The numerator calculates the power intercepted by the aperture, while the denominator represents the total radiated power integrated over a spherical surface centered at the feed.
\begin{align} \label{eq:eta_s}
\eta_s &= \frac{\iint_{A} P(r) \, dA}{\iint_{{\text{sphere}}} P(r) \, dA} \\
\text{where} \quad P(r) dA &= \frac{H}{r^3} \left(\frac{r_0^2 + r^2 - s^2}{2 r_0 r}\right)^{2q} dx dy
\end{align}

The illumination efficiency is related to how much of the target surface area is not illuminated, when considering the aperture of the feed and its distance, and is defined by \Cref{eq:eta_i}.
\begin{align} \label{eq:eta_i}
\eta_i &= \frac{1}{A} \frac{|\iint_{A} I(x,y) \, dA|^2}{\iint_{{A}} |I(x,y)|^2 \, dA} \\
\text{where} \quad I(x,y) &= \frac{H^{q_e}}{r^{1+q_e}} \left(\frac{r_0^2 + r^2 - s^2}{2 r_0 r}\right)^{q} 
\end{align}

The total aperture efficiency is defined by the product $\eta_a = \eta_s \cdot \eta_i$. Given this we optimized the aperture efficiency by performing sweeps over the physical configuration parameters of our experimental setup, considering $q = 5$ and a conventional value of $q_e = 1$ for the element pattern power factor. Specifically, we evaluate the efficiencies for a range of \SI{0}{\degree} to \SI{50}{\degree} for $\theta_0$, \SI{0.2}{\meter} to \SI{1.8}{\meter} for $H$, and $\pm\SI{15}{\centi\meter}$ for $y_0$, while keeping $x_0 = 0$, and considering the \ac{RIS} area of $\SI{18.46}{\centi\meter}^2$.

Following this, the optimal values were determined to be 
$H = \SI{33}{\centi\meter}$, $ \theta_0 = 35^{\circ}$, and $FBP = (0, \SI{-0.02}{\centi\meter}, 0)$. This setup resulted in about \SI{35}{\percent} efficiency due to the minimum possible H in the anechoic chamber. Higher efficiency would have required $H \approx \SI{16}{\centi\meter}$, but this was not possible due to physical limitations within the anechoic chamber and potential signal interference with the \ac{SoI}. Finally, the \ac{SoI} was shifted by \SI{15}{\degree} to prevent interference from the Tx antenna, and the Rx was placed directly below the \ac{RIS}, oriented towards the center of the \ac{SoI}.\\

\begin{figure}[t!]
    \centering
    \includegraphics[width=0.95\linewidth]{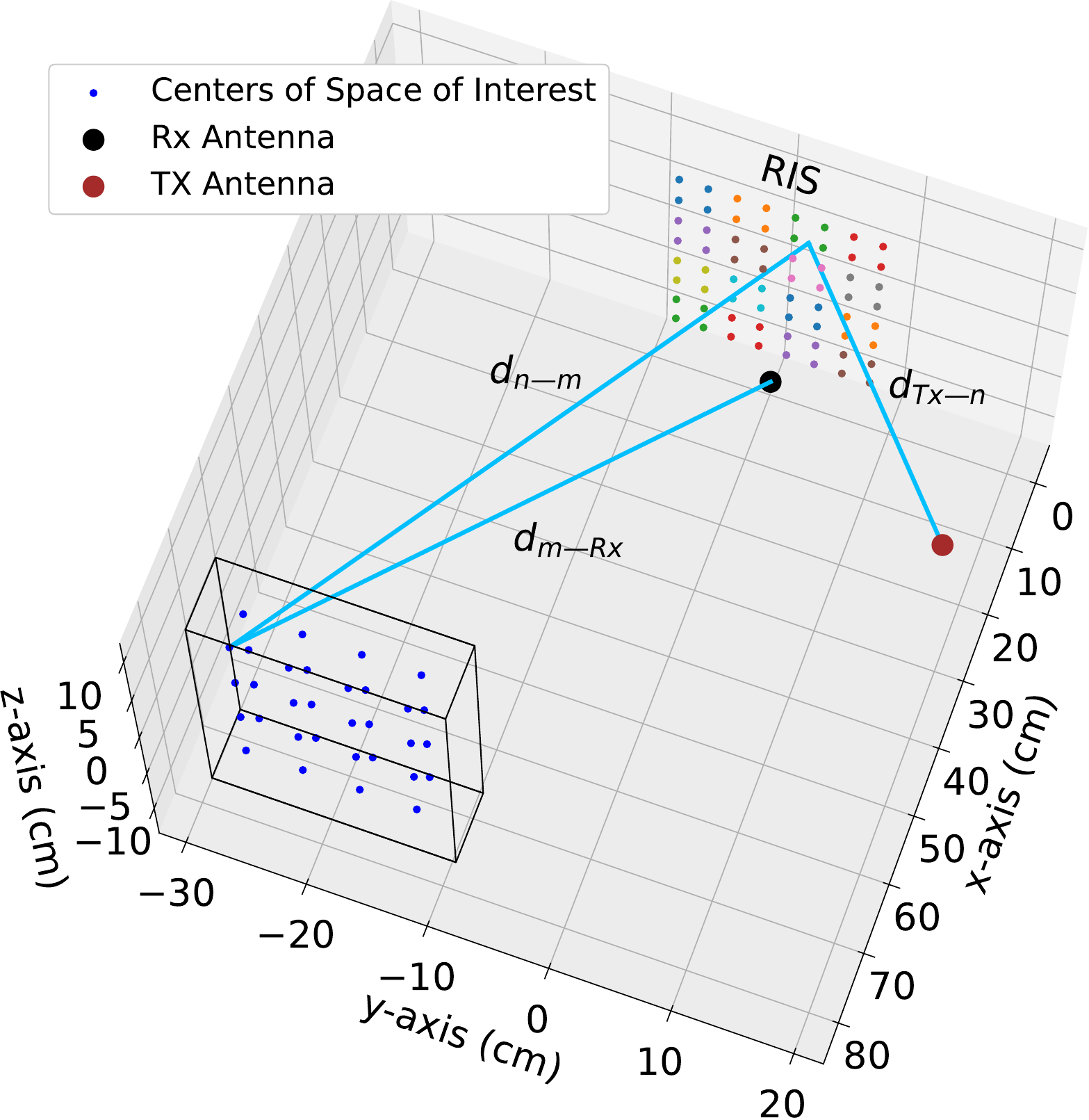}
    \caption{Top-view of modeled environment for our \ac{HGR} experiments.}
    \label{fig:esquema_3d_hgr}
\end{figure}

\noindent\textbf{Tx-to-Rx Propagation Model through RIS and SoI} 
The setup shown in \Cref{fig:esquema_3d_hgr} models the \ac{RIS}, \acf{SoI}, and Tx and Rx antennas, at positions given by the analysis explained in the previous section. The \ac{RIS} consists of $N \times N$ elements (where $N = 8$), divided into $L = 16$ square groups, each containing $N_l = 4$ elements. Each element of the \ac{RIS} is controlled by a single PIN diode, which can be in two states, ON or OFF. Therefore, each element can switch between two states, $s_i$, with $i \in \{1, 2\}$ and $N_a = 2$). For simplicity of control, it is assumed that all elements within the same group $l$ are in the same state $s_l$. The \ac{SoI}, located in front of the \ac{RIS}, is divided into $M = 32$ equal cuboids. Each cuboid is characterized by a reflection coefficient $\eta_m$, which is zero when no object is present and non-zero otherwise.

The most significant components are the multiple reflections from the transmitter to \ac{RIS} and the subsequent propagation of the signal to the \ac{SoI}, where further reflections occur off the object within the \ac{SoI}, ultimately captured by the Rx antenna. Thus, considering the states of all groups as the vector $\mathbf{s}_L$, the signal at the receiver can be expressed as:
\begin{equation}
y_{rx}= \sum _{m\in [1,M]}\sum _{l\in [1,L]}\sum _{n\in \mathcal N_{l}} h_{n,m}(s_{l},\eta _{m}) {\cdot P_{t} \cdot x}
\end{equation}
where $P_t$ represents the Tx Power, $x$ is a unit baseband with frequency $f_c$ and $h$ represents the channel gain. Based on \cite{goldsmith2005wireless}, $h$ is represented by:
\begin{align} 
\phi_{n,m} &= e^{-j 2 \pi ({d_{nTx}}+{d_{nm}}+d_{mRx}) / \lambda } \nonumber \\
\quad
h_{n,m}(s_{l},\eta _{m}) &= \frac{ \lambda \cdot r_{n,m}(s_l) \cdot \eta _{m} \cdot \sqrt{gT \cdot gR}}{ 4 \pi \cdot d_{nTx} \cdot d_{nm} \cdot d_{mRx}} \cdot \phi_{n,m}  \label{hd}
\end{align}

Here, $\lambda$ is the wavelength of the signal, and $gT$ and $gR$ denote the gains of the transmitter and the receiver, respectively. The distance from the Tx antenna to the \textit{n-th} \ac{RIS} element is given by $d_{nTx}$, $d_{nm}$ represents the distance from the \textit{n-th} RIS element in the \textit{l-th} group to the \textit{m-th} spatial block, and $d_{mRx}$ is the distance from the \textit{m-th} spatial block to the Rx antenna. Finally, $r_{n,m}$ are the reflection coefficients of the \ac{RIS} elements in a given state $s_i$, and $\eta_m$ are the reflection coefficients of each cuboid $m$ in the \ac{SoI}.

This model demonstrates that the signal at the receiver is a function of the known state of the \ac{RIS} elements, and unknown conditions of the \ac{SoI}. The following section explains how we now manipulate these states in order to relate the received signal to these unknown conditions to implement sensing.

\subsection{RIS Configuration Sequences for Sensing} 
\label{sub:risconfig}
Given our propagation model, we now present the method to determine a \ac{RIS} configuration sequence that optimizes the information extracted for the medium. The method protocol operates in a periodic time \textit{frame}, where each \textit{frame} has a duration $\delta$. Instead of changing the configurations of all \ac{RIS} elements simultaneously, each \ac{RIS} element sequentially transitions between states $s_1$ and $s_{Na}$ within each \textit{frame}.

For the $l$-th group of elements, the duration in each state is represented by the vector $\tilde{t}_l = (\tilde{t}_{l,1}, \ldots, \tilde{t}_{l,Na})^T$, with $\sum_{i=1}^{Na} \tilde{t}_{l,i} = \delta$. The complete matrix describing the time configuration $T$,  with dimensions $K \times (L N_a)$ where $K = 10$ frames, is defined as:
\begin{equation}  \label{T}  
T = \begin{pmatrix}
\mathbf{t_1} & \mathbf{t_2} & \cdots & \mathbf{t_K}
\end{pmatrix}^T 
\end{equation}
where ${t_k} = [t_{1,1}\ t_{1,Na}\ \ldots t_{L,1}\ t_{L,Na}]$. In conclusion, the received signal can be computed as: 
\begin{equation} \label{y_refl}
    y_{rx} = P_t \cdot x \cdot T \cdot A \cdot \eta, \ \ \ \text{with } \Gamma = T \cdot A  
\end{equation}

where matrix A represents the channel gain, with elements denoted by \(\alpha\). Each element represents the channel gain of the \ac{RF} paths from Tx to Rx through the $L$ \ac{RIS} groups in the different states and the M spatial blocks, resulting in a dimension of $LN_a \times M$:
\begin{align} \label{22}
 A &= [\mathbf{\alpha_1}\ \mathbf{\alpha_{2}} \dots \mathbf{\alpha_M} ] \nonumber \\
\quad
\mathbf{\alpha_m} &= [\alpha_{m,1,1} \dots \alpha_{m,1,Na} \dots \alpha_{m,L,1} \dots \alpha_{m,L,Na}]^T \nonumber \\
\quad
\alpha_{m,l,i} &= \sum_{n \in N_l} \frac{ \lambda \cdot r_{n,m}(s_l) \cdot \eta_{m} \cdot \sqrt{gT \cdot gR}}{ 4 \pi \cdot d_{nTx} \cdot d_{nm} \cdot d_{mRx}} \cdot \phi_{n,m}
\end{align}

Thus, the characterization of the \ac{SoI}, i.e., our \ac{RIS}-base \ac{RF} sensing, is based on the received signal $y_{rx}$, which can be reconstructed as:
\begin{equation}
    y_{rx} = \Gamma \cdot \eta + z
\end{equation}
where $\Gamma \in \mathbf{C}^{K \times M}$ represents the measurement matrix, $z \in \mathbf{C}^K$ denotes additive noise, and $\eta \in \mathbf{C}^M$. 

In other words, for a measured signal $y_{rx}$, and a known sequence of \ac{RIS} configuration states $T$, together with the channel gain of all unit cells, jointly represented by $A$, then we can derive the reflection coefficients $\eta_m$ of each cuboid in the \ac{SoI}. Thus, sensing of the space is implemented as each reflection coefficient $\eta_m$ will vary with the conditions (e.g., human posture or gesture) of the space.

However, the sequence of \ac{RIS} configuration states $\Gamma$ should be designed such that the most information can be derived from the space. In simple terms, the signal paths between each unit cell, the \ac{SoI}, and the Tx antenna, should carry the least correlated amount of information possible.

We achieve this optimization of $\Gamma$ by adapting the method presented by Hu \textit{et al.} \cite{9133157} to our two-state \ac{RIS}. Specifically, we adapt their \ac{FCAO} algorithm, which is proposed for dynamic control of \ac{RIS} elements, to minimize the mutual coherence of the $\Gamma$ matrix elements. We then employ the resulting \ac{RIS} control patterns in our \ac{HGR} experiments, demonstrating how the \ac{FCAO} generated patterns extract richer information from the \ac{RF} medium relative to randomly applied patterns.

\subsection{Unit Cell and RIS Tile Implementation}
\label{sub:risimplement}

\begin{figure}
\centering
\subfloat[top layer]{\includegraphics[width=0.485\columnwidth]{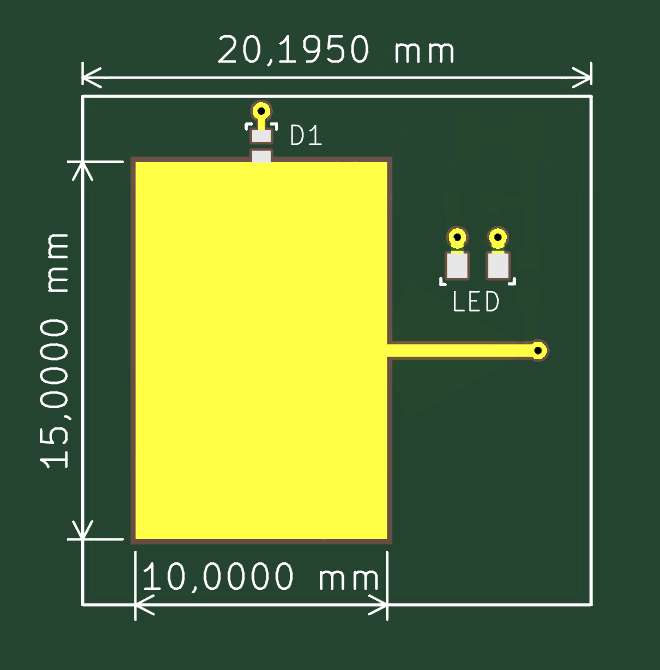}%
\label{fig:unitcelltop}}
\hfil
\subfloat[bottom layer]{\includegraphics[width=0.485\columnwidth]{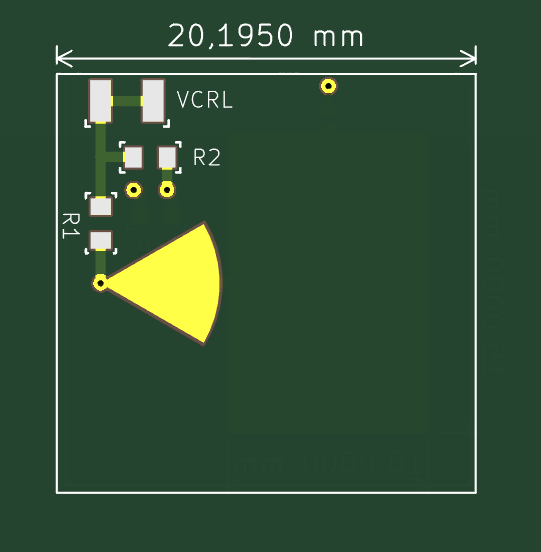}%
\label{fig:unitcellbottom}}
\caption{\SI{6.5}{\giga\hertz} Unit cell design with PIN diode control}
\label{fig:unitcell}
\end{figure}

In this section we present our design for a 1-bit PIN diode based unit cell, and the 64-element \ac{RIS} based on this element. All electromagnetic design and simulation was performed through CST Studio. Our objective was to design and test a \ac{RIS} tile operating the WiFi-6E range and validate its capacity to manipulate the electromagnetic medium for a given range of frequencies centered on \SI{6.5}{\giga\hertz}, providing the hardware necessary for our \ac{RF} based \ac{HGR} approach.\\

\noindent\textbf{6.5\,GHz PIN Diode Based Unit Cell Design}\label{sub:unitcell} The unit cells were designed to be tuned to for a center frequency of \SI{6.5}{\giga\hertz}, corresponding to the central frequency of the WiFi-6E band. Its' top and bottom layer designs are shown in \Cref{fig:unitcell}. The top layer contains a rectangular patch of \SI{10}{\milli\meter} by \SI{15}{\milli\meter}. The total enclosing perimeter of the design is a square with a side of \SI{23}{\milli\meter}. Placing two cells side by side provides a separation of half a wavelength. Also shown are the contacts for the D1 PIN diode used to control the phase response. Due to its placement, the unit cell has a vertical polarization. The control voltage for the PIN diode is applied on the bottom layer (\Cref{fig:unitcellbottom}), at \emph{VCRL}. A radial stub is used to decouple the RF path from the DC path.

The PIN diode is an SMP1331-079LF, and the control voltage applied at the \emph{ON} state is of \SI{0.8}{\volt}. Given the \SI{220}{\ohm} value for \emph{R1} and the value of \SI{3.3}{\volt} applied at \emph{VCRL}, a unit cell in an \emph{ON} state draws \SI{15}{\milli\ampere}. Thus, if all PIN diodes are conducting, the \ac{RIS} consumes \SI{3.16}{\watt}, but the majority of control patterns only activate approximately half the unit cells. Resistor \emph{R2} regulates the current for the status LED, which exists to allow for easy interpretation and validation of the applied control pattern throughout the \ac{RIS}. The LEDs can be disabled globally, but each draws \SI{1}{\milli\ampere} if active.

The PCB has a composite substrate, where the top layer is a \SI{4}{\milli\meter} F4B substrate, and the bottom layer is a \SI{0.5}{\milli\meter} FR4 substrate. The cost of a tile totaled approximately 85\texteuro~for fabrication and approximately 60\texteuro~for components. The PIN diodes alone are around \SI{50}{\percent} of this cost.\\

\begin{figure}
\centering
\includegraphics[width=0.90\columnwidth]{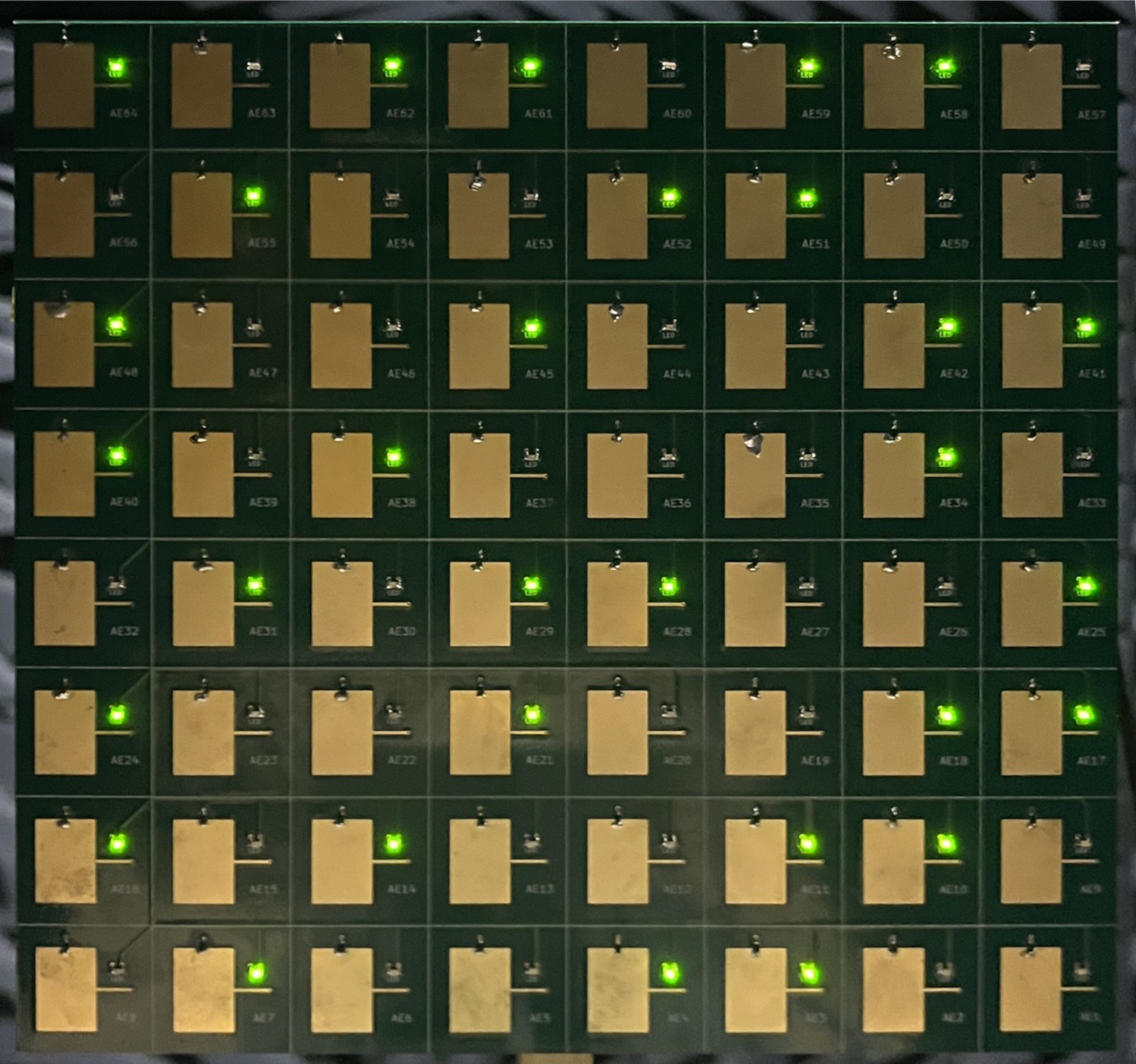}
\caption{64 element RIS tuned for \SI{6.5}{\giga\hertz}, with 1-bit phase control (\SI{18.46}{\centi\meter} by \SI{18.46}{\centi\meter}). The status LEDs show some unit cells at the \emph{ON} state.}
\label{fig:ristile}
\end{figure}

\noindent\textbf{8x8 RIS Tile with 6.5\,GHz Unit Cell}
The fabricated \ac{RIS} tile is shown in \Cref{fig:ristile}. It is composed of 64 unit cells designed as shown in \Cref{fig:unitcell}, resulting in a square design with a side of \SI{18.46}{\centi\meter}. In \Cref{fig:ristile} some of the LEDs, one per unit cell, are active, indicating that the respective unit cell is at its \emph{ON} state, which should ideally produce a \SI{180}{\degree} of phase shift for the signal reflection. A total of 64 signals are needed to control the entire tile. 

\begin{figure}
\centering
\includegraphics[width=0.90\columnwidth]{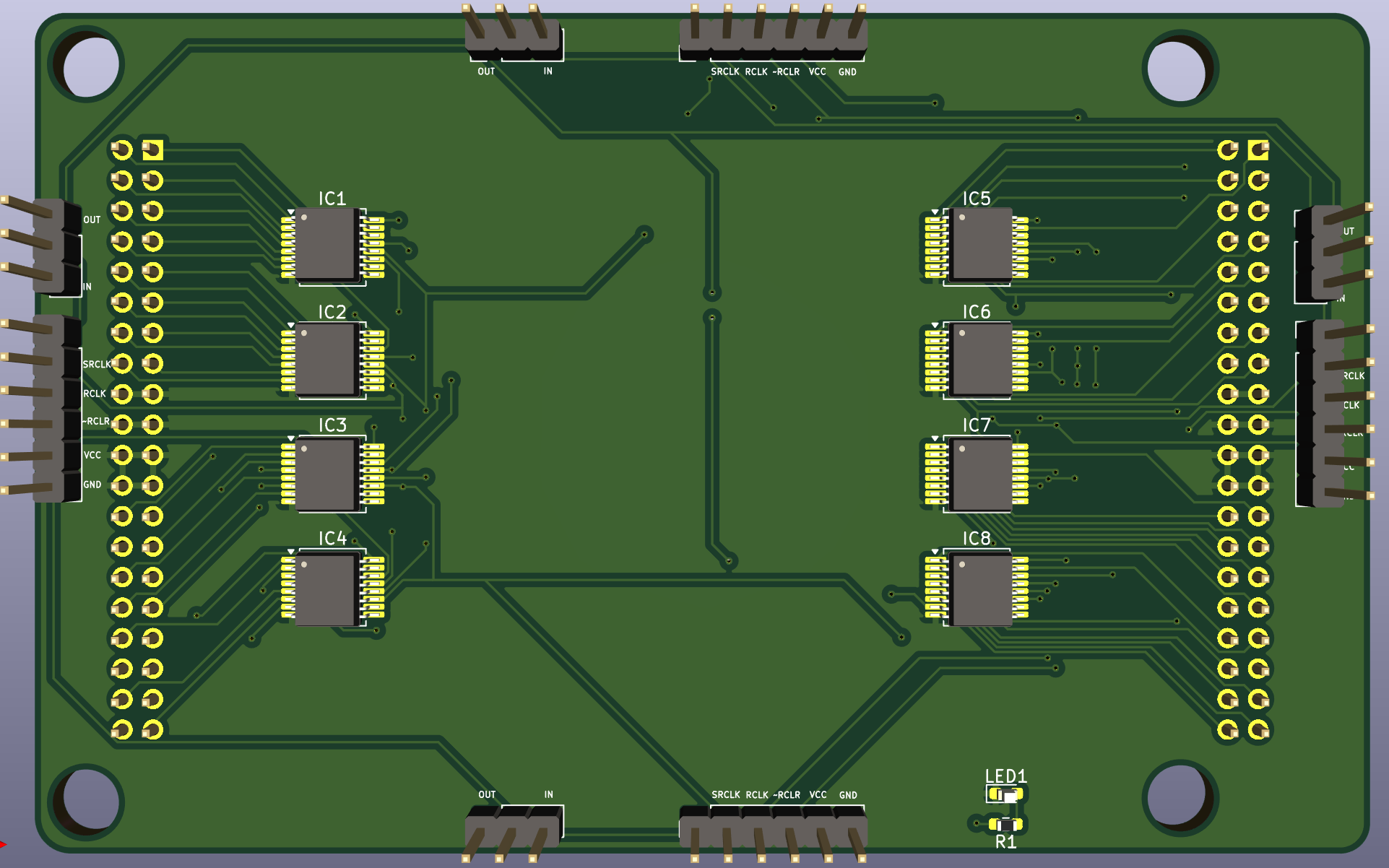}
\caption{Top layer of control board hosting shift registers to hold the configuration pattern. The bottom layer contains two 40 male pin-header connectors to interface with the tile.}
\label{fig:controlboard}
\end{figure}

The tile allows for easy scalability without compromising ideal beamforming conditions. It was purposefully designed such that when physically placing two tiles side by side, with direct contact between the edges of the PCBs, the spacing between elements at the periphery of both tiles equals then spacing within a single tile. A multi-tile \ac{RIS} can be assembled by resorting to a mounting structure that secures each tile via its bottom layer, as opposed to any frame based mounting which would physically displace the relative tile positions. We achieve this by assembling two SMD 40-pin female pin header connectors on the bottom layer of the tile. These connectors provide the physical mounting support, specifically by interfacing with a smaller control board, shown in \Cref{fig:controlboard}. 

The control board is \SI{11}{\centi\meter} by \SI{7}{\centi\meter}, and hosts eight shift-registers. Namely, eight 74HC595PW ICs, which are 8-bit serial-in parallel-out shift registers. We have connected these in a daisy chain, and to set these shift registers, the control board contains an SPI interface to an ESP32-S3-DevKitC-1. The ESP32 sends 64 bits to the shift-registers, and issues the a \emph{latch} signal which changes all unit cell configurations simultaneously. This synchronization prevents undesired intermediate states which would occur if the values were sent directly to the unit cells as their were being shifted into the tile.

We opt to implement this control board separately from the \ac{RIS}, instead of assembling the shift-registers on its bottom layer, to mitigate any effects on beamforming performance and to simplify the \ac{RIS} by homogenizing it, which also simplifies scalability. This separation also allows using the same control board for other \ac{RIS} designs, e.g., for different unit cell designs. The total cost of the control board is approximately 40\texteuro. 

Finally, besides allowing for the rear mounting which facilitates the physical scaling of the number of tiles, the four interfaces at the control board's periphery can be interconnected to control additional unit cells. For example, a single ESP32 can command the resulting 256 elements of 4 tiles as a single larger \ac{RIS}, or four parallel daisy chains can allow for a more explicit tile-level control.

\section{Experimental Evaluation}
\label{sec:experiments}

We conducted several experimental campaigns. Firstly, we validate the \ac{RIS} design, which includes characterizing the magnitude and phase response of the fabricated unit cell versus simulations (\Cref{sub:unitcellcharacterization}), and measuring the beamforming capability of the \ac{RIS} as a whole (\Cref{sub:rischaracterization}). Once the correct functioning of the \ac{RIS} was established, we conducted \ac{HGR} related experiments, namely a preliminary trial to determine the viability of hand gesture recognition with our setup (\Cref{sub:hgrwood}), followed by a more extensive data gathering to perform the \ac{CNN} based classification of gestures (\Cref{sub:hgrhuman}).\\

\noindent\textbf{Automated Data Aquisition} 
To ensure fast and consistent execution of all necessary experimental run configuration, data acquisition, and post-processing tasks across experiments, we developed software to interface with our laboratory instruments. Specifically, through a laptop controlling the process, we synchronized processes such as \ac{RIS} configuration, adjusting relative orientations of Tx and Rx, and measuring $S_{21}$ data. After setting a particular physical and digital configuration of the system, we resort to a Python-based \ac{VISA} library to record the corresponding $S_{21}$ measurements from the \ac{VNA}. Specifically, we resort to a portable Keysight Fieldfox N9914A \ac{VNA}, to facilitate the assembly of this data-gathering setup and self-contained experimental setups within our anechoic chamber. The design files and control software for these components are publicly accessible\footnote{F. Ribeiro and M. Oliveira, ``SpecRF-Posture". GitHub, Apr. 03, 2024. \url{https://github.com/franciscombr/SpecRF-Posture}}. The following experiments (except for the unit cell characterization via a waveguide) rely on minor variants of this automated flow to gather all measurements.

\subsection{Unit Cell Characterization in Waveguide}
\label{sub:unitcellcharacterization}


In this first experiment, we measured the true response of the fabricated unit cell, versus the expected response as given by CST simulation. This verification then determined the experimental parameters for the \ac{RIS} characterization experiments (\Cref{sub:rischaracterization}), namely the frequency with the best expected steering response, and also influenced the analysis of the \ac{HGR} results (\Cref{sub:hgrhuman,sub:hgrwood}).\\ 

\begin{figure}
\centering
\includegraphics[width=0.85\columnwidth]{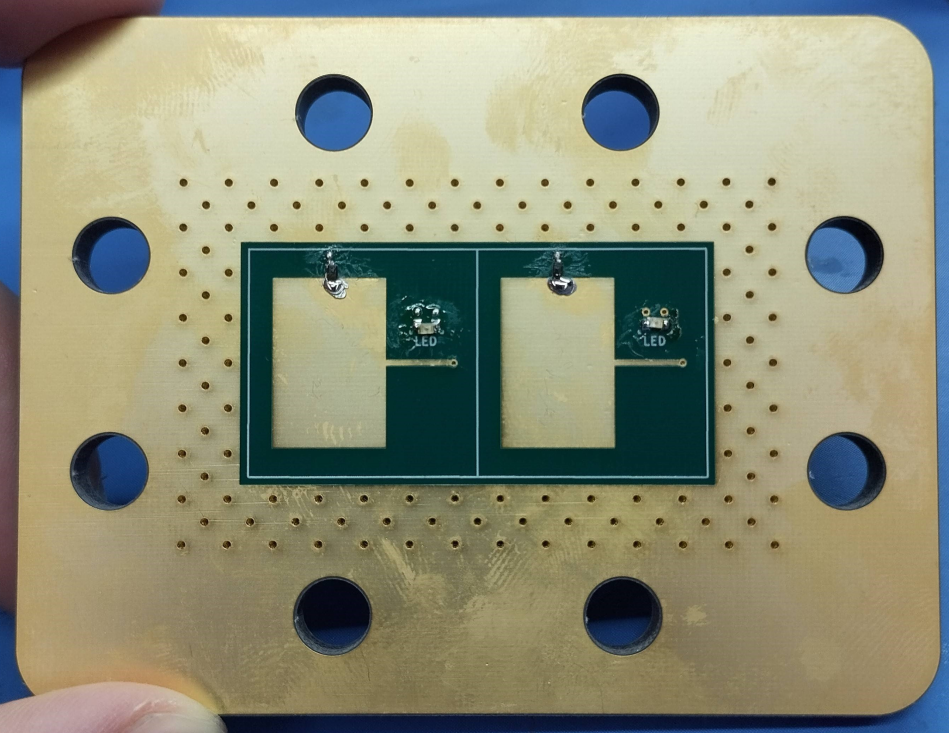}
\caption{PCB with two unit cells for validation of frequency and magnitude/phase response in a WR159 waveguide}
\label{fig:waveguidepin}
\end{figure}

\noindent{\textbf{Experimental Setup}}
\Cref{fig:waveguidepin} shows the fabricated PCB for use with a WR159 waveguide, with two unit cells spaced identically to the \ac{RIS}. To measure the response, we applied the same control to both unit cells via an external voltage source. We measured the S11 parameters of the unit cell using a Keysight N5224B, a \ac{VNA} which operates up to \SI{43}{\giga\hertz}. The simulations considered a Floquet port setup and periodic boundary condition in the 3D electromagnetic simulation software CST, and range from \SI{5.8}{\giga\hertz} to \SI{7.2}{\giga\hertz}, given the intended center frequency of \SI{6.5}{\giga\hertz}. The measurements cover a lower range between \SI{5}{\giga\hertz} to \SI{7}{\giga\hertz}, given the response we observed.\\

\begin{figure}
\centering
\subfloat{
\includegraphics{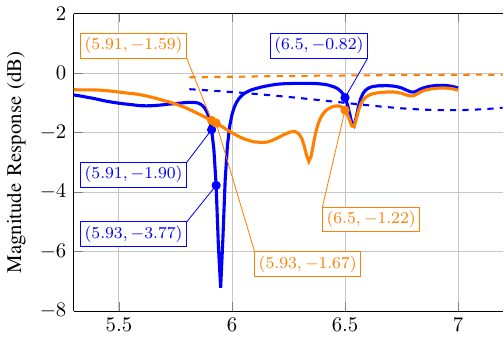}
\label{fig:unitcellfreq}}
\hfil
\subfloat{
\includegraphics{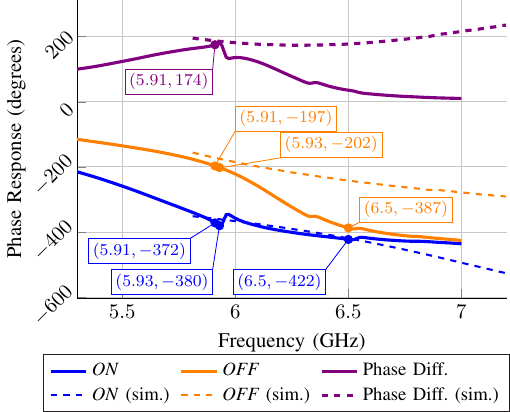}
\label{fig:unitcellphase}}
\caption{Magnitude and phase response of unit cell measured in \ac{VNA}, for both \emph{ON} and \emph{OFF} states. A phase difference closest to \SI{180}{\degree} is achieved for \SI{5.93}{\giga\hertz}, but a better magnitude response is present at \SI{5.91}{\giga\hertz}.}
\label{fig:unitcellresponse}
\end{figure}

\noindent{\textbf{Experimental Results}}
\Cref{fig:unitcellresponse} shows the resulting S11 measurements and respective simulations, for the \emph{ON} and \emph{OFF} states. \Cref{fig:unitcellfreq} shows the magnitude response. At \SI{6.5}{\giga\hertz} the response for both states is approximately equal, which is a desired behavior since the reflected power should not depend on the applied control, to avoid any distortions to the desired beamforming. However, the phase difference between states is only \SI{35}{\degree}, whereas the simulated response was \SI{180}{\degree}.

Instead, the greatest phase difference of \SI{178}{\degree} is observed for \SI{5.93}{\giga\hertz}. However, at this frequency the \emph{ON} state experiences a \SI{-3.77}{\decibel} attenuation. Since the magnitude response of the \emph{OFF} state at this frequency is \SI{-1.67}{\decibel}, the disparity in magnitude response between states could 
affect beamforming performance. A trade-off between a balanced magnitude response and phase difference occurs for \SI{5.91}{\giga\hertz}, where a \SI{174}{\degree} phase difference is attained for a more approximate magnitude response between states, and where the response for either state is not lesser than \SI{-3}{\decibel}.

This deviation may be due to the LED which was not simulated, or to the tolerances of the PIN diode, whose datasheet reports a typical \SI{0.18}{\pico\farad} capacitance up to a maximum of \SI{0.35}{\pico\farad}. However, we note that the magnitude response was extremely sensitive to the alignment between the PCB and the waveguide, with the phase response less so. Additionally, it is known that the waveguide environment imposes an oblique wave incidence condition (of approximately \SI{38}{\degree} in this case), unlike the normal incidence obtained in free space conditions.

\noindent{\textbf{Conclusion}}
It is difficult to determine precisely the source of the deviations from the design, or the true response that is obtained in free space. In the absence of a better reference, for the following tests of the \ac{RIS}, we consider \SI{5.91}{\giga\hertz} as the central frequency for computing beamforming control patterns.

\begin{figure*}[!t]
\centering
\subfloat[relative TX to RIS positioning, with TX feed point positioned \SI{400}{\milli\meter} away from center of RIS and a \SI{35}{\degree} angle]{\includegraphics[height=0.282\textwidth]{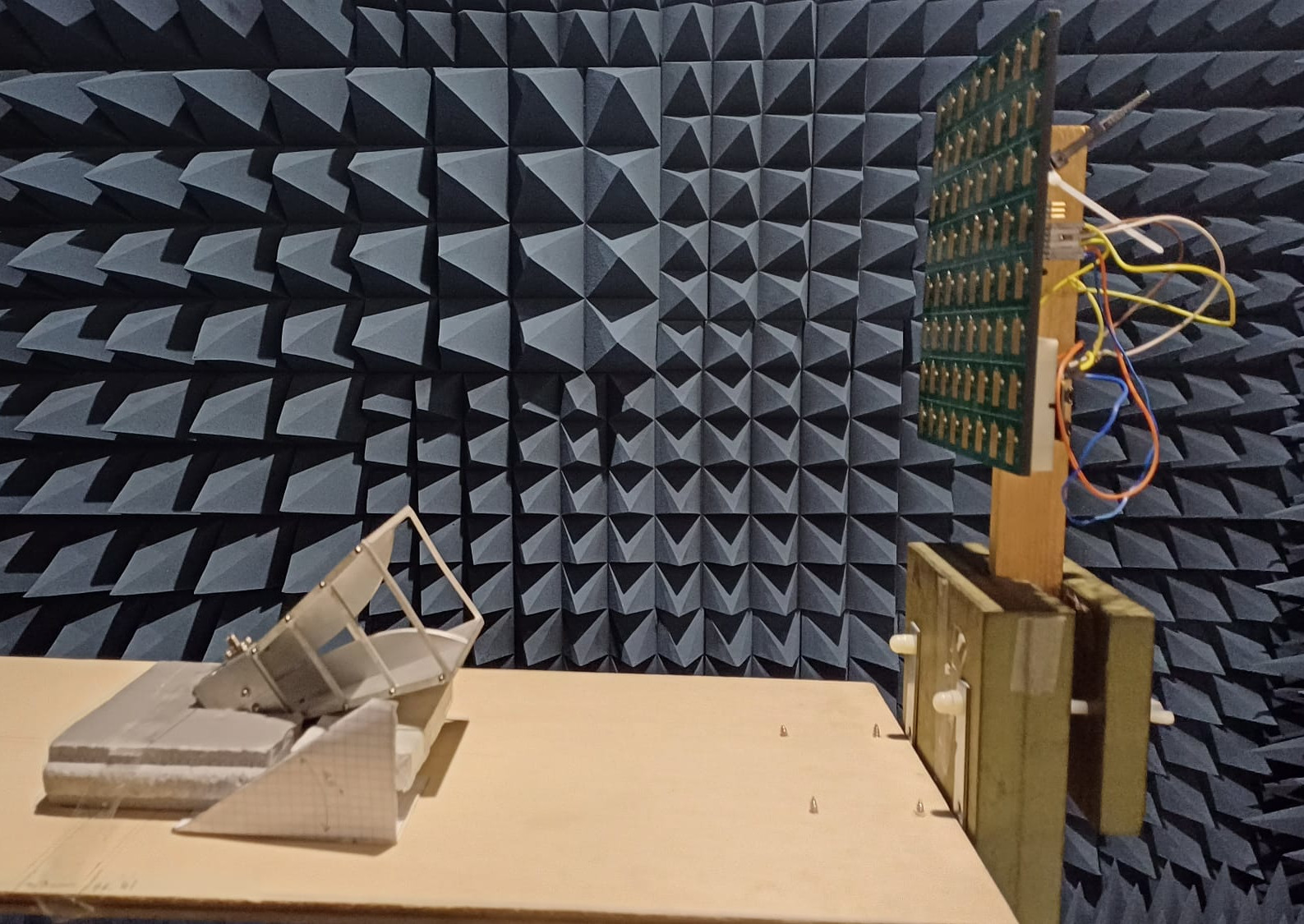}%
\label{fig:chamber1}}
\hfil
\subfloat[RX position relative to RIS mounted on rotor]{\includegraphics[height=0.282\textwidth]{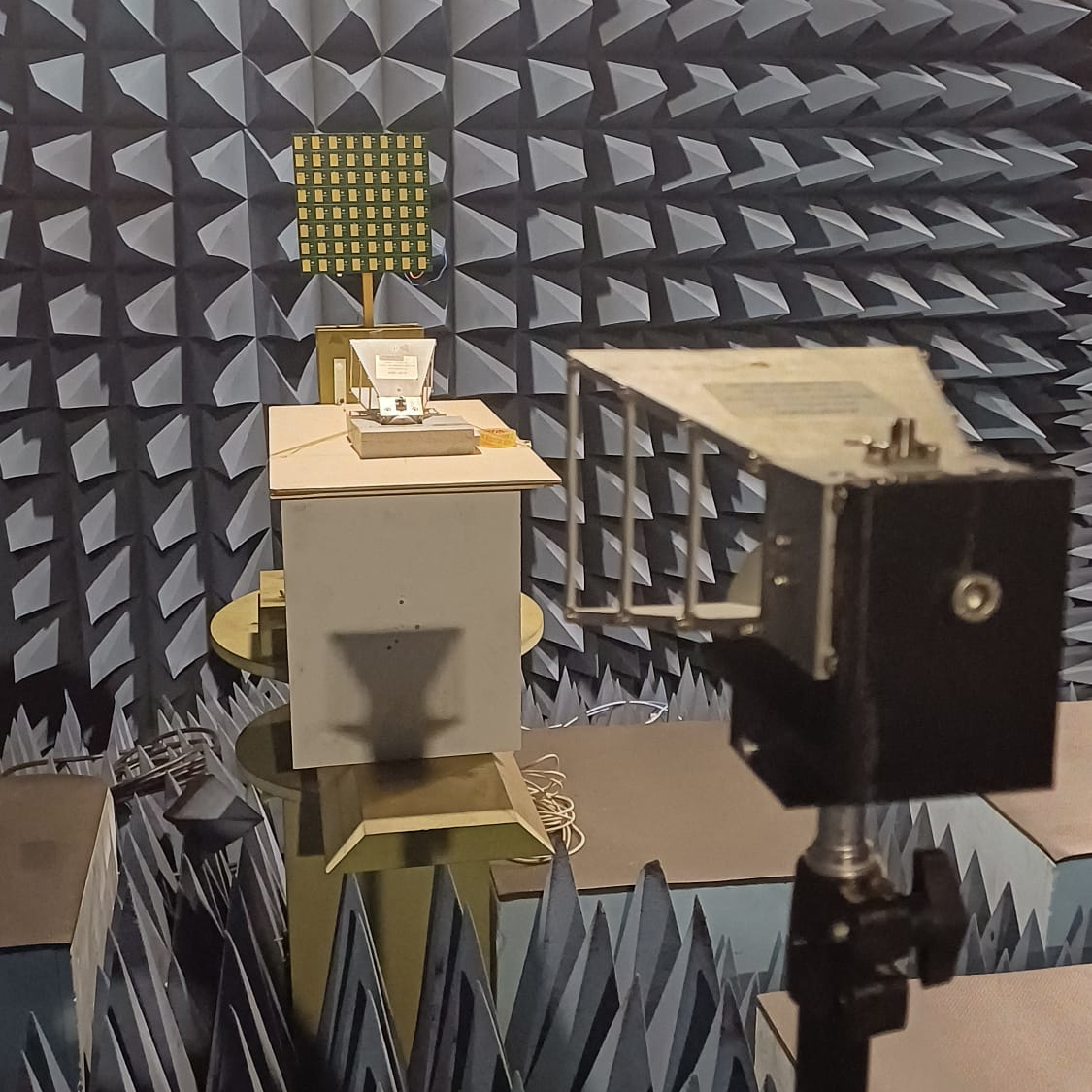}%
\label{fig:chamber2}}
\hfil
\subfloat[rotor positioned at \SI{-60}{\degree} relative to its rotation axis, with relative position between TX and RIS maintained]{\includegraphics[height=0.282\textwidth]{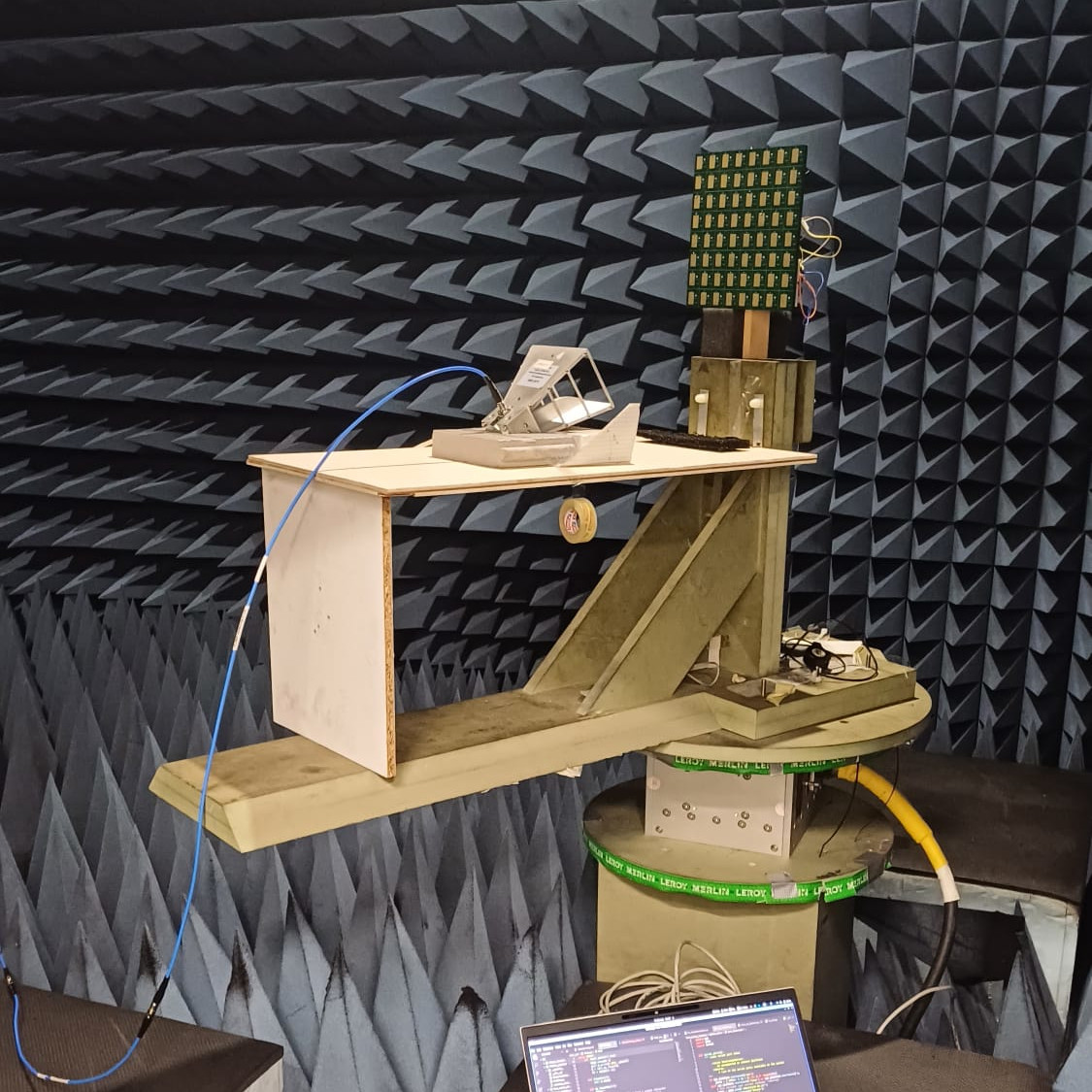}%
\label{fig:chamber3}}
\caption{Setup for \ac{RIS} data gathering with RIS and TX mounted on a rotational platform, RX antenna on a fixed support. The VNA (not shown) and rotor control are both connected to a latop to drive automated data gathering.}
\label{fig:chambersetup}
\end{figure*}

\subsection{RIS Characterization in Anechoic Chamber}\label{sub:rischaracterization}

To validate and characterize the \ac{RIS}, we measure its beamforming capabilities for several combinations of digital beamsteering and relative orientation between the \ac{RIS} and a receiver. Following we explain the physical setup in an anechoic chamber of approximately $\SI{7}{\meter}\times\SI{7}{\meter}\times\SI{3}{\meter}$, and the theory behind the generated \ac{RIS} control.\\ 

\noindent{\textbf{Experimental Setup}}
\Cref{fig:chambersetup} shows our setup in an anechoic chamber. We employed the LB-20180-SF horn antenna model identified in \Cref{sub:modelsetup} for both the Rx and Tx, and the mentioned portable \ac{VNA} to measure the $S_{21}$ parameters. We resort to our automated data gathering setup, using it to control the \ac{RIS} digital beamforming, the position of a rotor assembly, and to gather data from the \ac{VNA}. 

We consider that the center of the coordinate system is at the center of the \ac{RIS}, that the XY plane is the broadside of the \ac{RIS}, and that the direction of the X axis is downwards. The \ac{RIS} is mounted on a rotating platform, and its X axis is aligned with the rotational axis of this platform. The feed horn was placed on a platform below the \ac{RIS}, displaced by \SI{25.5}{\centi\meter} on its X axis, and \SI{36.5}{\centi\meter} on its Z axis, resulting in a distance of \SI{40}{\centi\meter} to the feed point phase center. This distance was chosen as a compromise between the ideal distance of \SI{20}{\centi\meter}, which would provide an edge tapper of \SI{-9}{\decibel} (as per \Cref{sub:modelsetup}), and a setup that would reduce the occlusion of the \ac{RIS} by the feed horn itself. 
The Rx antenna was placed \SI{170}{\centi\meter} away from the \ac{RIS}, approximating far field conditions for the measured frequencies. 

The \ac{RIS} control board is attached behind the PCB, containing also the ESP32 micro-controller powered by a power bank, which receives the pattern to apply to the \ac{RIS}  from a laptop via WiFi (on a \SI{2.4}{\giga\hertz} link). This laptop computes the configuration based on the desired steering, according to the following equations.
\begin{equation} \label{eq1}
R = \sqrt{(x_c - x_{ij})^2 + (y_c - y_{ij})^2 + z_c^2}
\end{equation}
\begin{equation} \label{eq2}
\phi_{ij} = k \left( R - \sin(\theta) x_{ij} \cos(\phi) + y_{ij} \sin(\phi) \right)
\end{equation}

The coordinates of the feed antenna are given by ($x_c, y_c, z_c$), which for this \ac{RIS} characterization setup are $(d\sin35^\circ, 0, d\cos35^\circ)$, given the distance $d = $ \SI{400}{\milli\meter} and \SI{35}\degree angle of the feed horn towards the \ac{RIS}. The coordinates of the center of each unit cell, \emph{i, j}, considering that the center of the coordinate system is at the center of the \ac{RIS}, are given by $(x_{ij}, y_{ij}) = (i C_d \minus \frac{H}{2}, j C_d \minus \frac{W}{2})$, where $C_d$ is the dimension of the unit cell, \SI{23}{\milli\meter}, and $H$ and $W$ are the height and width of the \ac{RIS}, both \SI{18.4}{\centi\meter}.

The beam direction is given by $\theta$ and $\phi$, $R$ is thus the Euclidean distance between each unit cell and the feed horn, $k = {\tfrac {2\pi }{\lambda}}$ is the wave vector. For these calculations, we considered the wavelength for \SI{5.91}{\giga\hertz}. Finally, given the two control levels available through the PIN diodes, we quantize the resulting ideal phase responses, $\phi_{ij}$, to either zero (for $0^\circ < \phi_{ij} < 180^\circ$) or one (for $180^\circ \le \phi_{ij} \le 360^\circ$).

With this setup, and considering a frequency range from \SI{5.3}{\giga\hertz} to \SI{6.5}{\giga\hertz} (the limit of the \ac{VNA}), we measured the $S_{21}$ parameters for \ac{RIS} steering configurations of \SI{0}{\degree}, $\pm$\SI{20}{\degree} and $\pm$\SI{40}{\degree}, for a range of rotor orientations of $\pm$\SI{60}{\degree} with a step of \SI{2}{\degree}.\\

\begin{figure}
    \centering
    \includegraphics{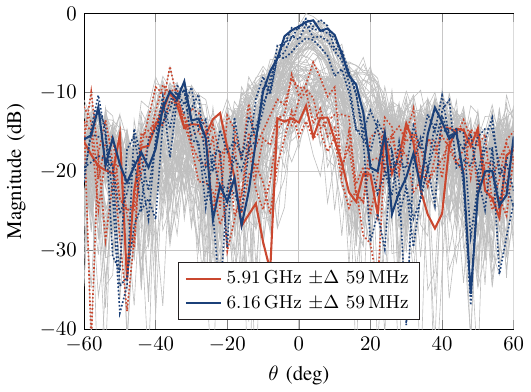}
    \vspace{-12pt}
    \caption{\ac{RIS} radiation patterns for a beamforming configuration of $\theta = 0^\circ$ and $\phi = 0^\circ$ (i.e, steering towards the normal vector of the \ac{RIS} plane) and an angle range of $\pm$\SI{60}{\degree}, with \SI{5.91}{\giga\hertz} and \SI{6.16}{\giga\hertz} highlighted.}
\label{fig:radiationSteer0}
\end{figure}

\noindent\textbf{Experimental Results}
We first retrieved the radiation pattern for \ac{RIS} configuration of $\theta = 0^\circ$, with a rotor orientation of \SI{0}{\degree} (i.e., forward-facing). Our \ac{VNA} resolution produced measurements for 200 frequencies, and \Cref{fig:radiationSteer0} shows the resulting radiation patterns, normalized to \SI{0}{\decibel}.

Plotted in light gray are the responses for a subset of all frequencies covering the measured range (specifically, from \SI{5}{\giga\hertz} to \SI{7}{\giga\hertz} with a step of \SI{5.9}{\mega\hertz}). We highlighted the previously chosen frequency of \SI{5.91}{\giga\hertz}, and four additional responses in a bandwidth of $\pm$\SI{59} 
{\mega\hertz}. Despite the expected best phase response at this frequency, there is no noticeable steering towards the desired direction of \SI{0}{\degree}. Instead, the shown $\pm$\SI{50}{\mega\hertz} bandwidth centered on \SI{6.16}{\giga\hertz} shows the most response to this steering configuration. 

We can attribute this to our measurement of the location of the feed point relative to the \ac{RIS} center, which is difficult to determine 
precisely in real-world measurements. The calculation of the phase patterns, shown in \Cref{eq1,eq2}, depends on these coordinates as well as the presumed central frequency. So, a likely deviation to the true value of ($x_c, y_c, z_c$) resulted in phase profiles more effective for \SI{6.16}{\giga\hertz} despite the value of \SI{5.91}{\giga\hertz} considered for the calculations.

\begin{figure}
    \centering
    \includegraphics{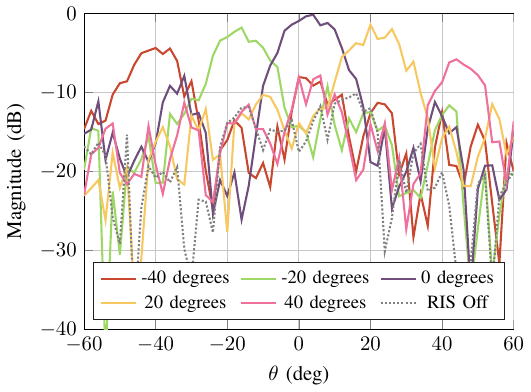}
    \vspace{-12pt}    
    \caption{RIS radiation patterns for six beamforming configurations, for \SI{6.16}{\giga\hertz}, showing an effective steering capability which compensates the rotor positioning. Also shown is the specular response (i.e., RIS off).}
\label{fig:radiationSteerAngles}
\end{figure}

We then measured the radiation patterns for the \ac{RIS} steering configurations of $\pm$\SI{20}{\degree} and $\pm$\SI{40}{\degree}. Given the previous conclusion, \Cref{fig:radiationSteerAngles} shows these profiles only for \SI{6.16}{\giga\hertz} where a good steering response is confirmed. The \ac{RIS} displays a symmetric behavior, without angle skew in regards to the steering direction, and with similar magnitude responses for opposing steering angles (e.g. $\pm$\SI{20}{\degree}). 

Despite the good steering response for this frequency, it may not be the best performance achievable by the design. According to the unit cell response for \SI{6.16}{\giga\hertz} seen in \Cref{fig:unitcell}, the phase difference between states is of approximately \SI{110}{\degree} and there is a difference of \SI{1.88}{\decibel} between states. That is, although the computed phase profiles generated a good steering response at this frequency, more accurate measurements of the feed point or adjustments to its position could result in a better response for \SI{5.91}{\giga\hertz}, exploiting a greater phase difference between states and therefore resulting in a more focused and power efficient beam. Finally, we also plot the radiation pattern measured with the \ac{RIS} disabled, i.e., acting as a purely specular reflector. This shows that there is little passive component received at the RX, for this frequency, and that the previous profiles are a the result of the beamsteering.\\

\noindent\textbf{Conclusion}
In summary, the \ac{RIS} correctly performs beamforming, achieving the best steering response for phase control patterns computed assuming a central frequency of \SI{5.91}{\giga\hertz}. Although a more pronounced steering response is observed for \SI{6.16}{\giga\hertz}, the \ac{RIS} does demonstrate steering behavior for a range of \SI{5.0}{\giga\hertz} to \SI{6.5}{\giga\hertz}. This means that, to a greater or lesser extent depending on the frequency, the electromagnetic medium is indeed influenced by the \ac{RIS} in this broader band. In other words, we can apply phase control patterns generated by the \ac{FCAO} based optimization approach, and measure the resulting S$_{21}$ for this range to drive our \ac{HGR} approach.

\subsection{RIS-based \acl{HGR} Feasibility Study} 
\label{sub:hgrhuman}

The objective of this first experiment was to validate not just the overall experimental setup, but mainly our methodology \acf{HGR}. Specifically, we aimed to prove that measurements of the $S_{21}$ parameters, considering the channel through the \ac{SoI}, reacted to the presence of a human hand to such a degree that different hand gestures could differentiated.\\

\noindent\textbf{Experimental Setup}
The experimental setup was assembled as we presented in  \Cref{sub:modelsetup}, within the same previously mentioned anechoic chamber, resorting to the same pair of horn antennas, i.e., two A-INFO LB-20180 pyramidal horns. Additionally, a ZX60-153LN-S+ low-noise amplifier was added to the setup, to amplify the signal of the Tx antenna.

\begin{figure}[!t]
\centering
\subfloat[side view of experimental setup using real hand (employed in \Cref{sub:hgrhuman})]{
\includegraphics[trim={0 30 0 210}, clip, width=1.00\linewidth]{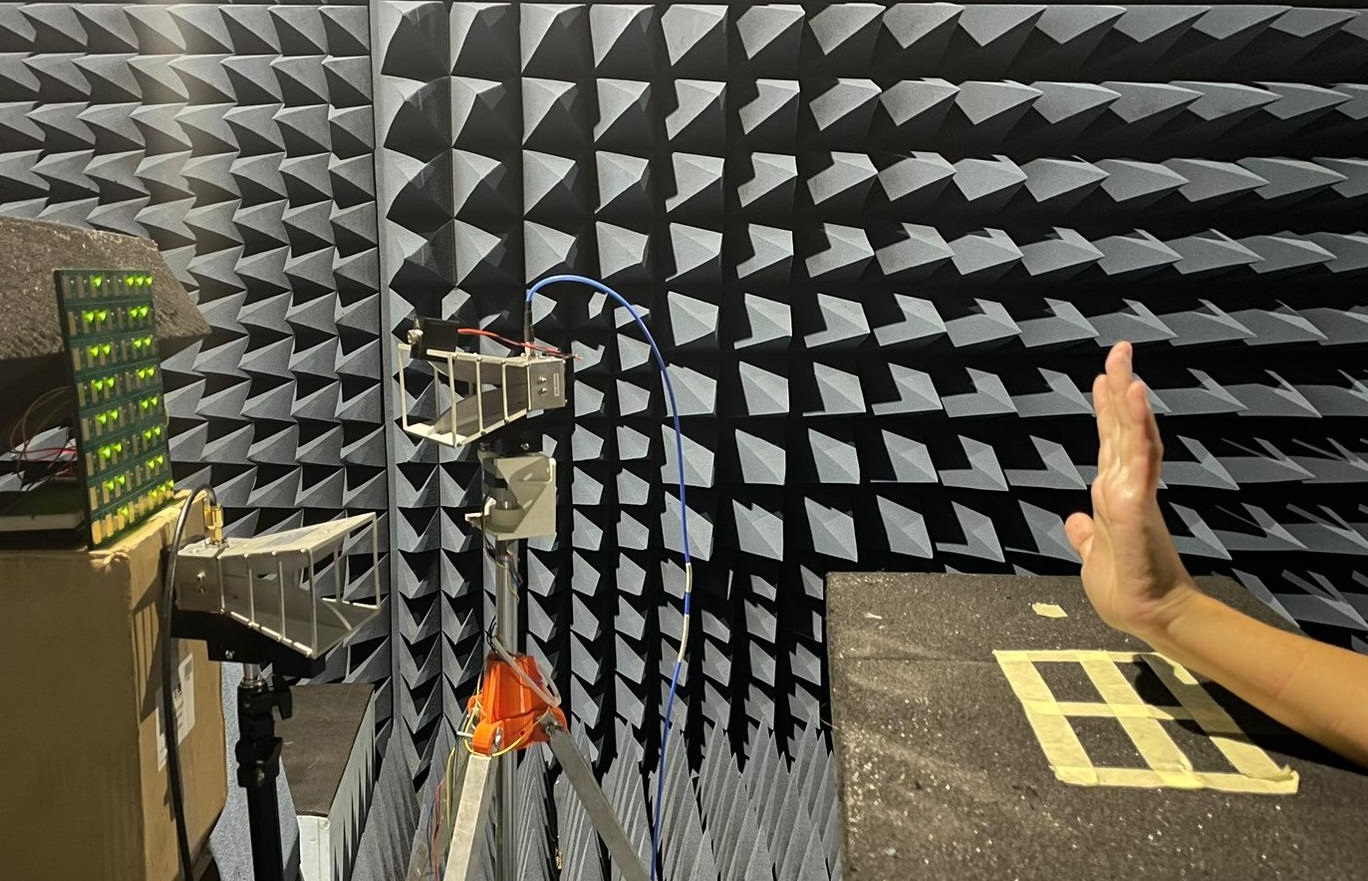}%
\label{fig:setup_ris_real}}

\subfloat[side view of experimental setup using model wooden hand (employed in \Cref{sub:hgrwood})]{
\includegraphics[trim={0 50 0 450}, clip, width=1.00\linewidth]{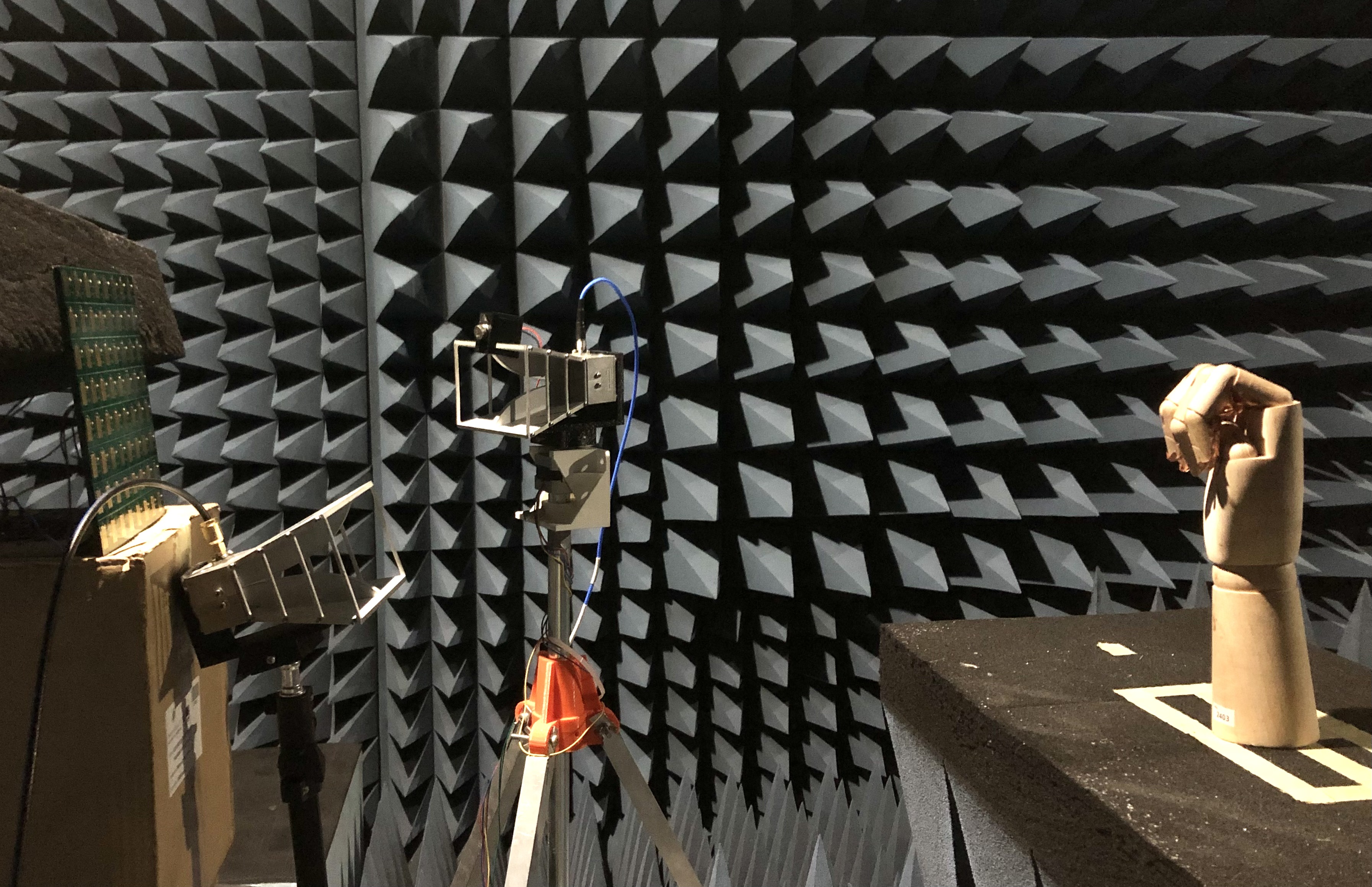}%
\label{fig:setup_ris_wooden}}
\caption{Experimental setup for HGR with (a) real hand (b) wooden hand, inside the anechoic chamber. Data was captured for three different hand gestures, plus an empty \ac{SoI}.}
\label{fig:setups_2}
\end{figure}

Following the model developed in \Cref{sub:modelsetup} (shown in \Cref{fig:esquema_3d_hgr}), \Cref{fig:setup_ris_real} illustrates the experimental setup for this experiment. Data was collected for three different hand gestures while the \ac{RIS} was disabled (\textit{off}), and for four distinct \ac{RIS} steering configuration, namely $5^{\circ}, 10^{\circ}, 15^{\circ}, 20^{\circ}$. Unlike other experiments, the presence of a human volunteer compromises the full automation of data gathering. Instead, for each of the three gestures, data acquisition involved manually saving the $S_{21}$ parameters from the \ac{VNA} after directly sending the \ac{RIS} configuration and assuring the correct positioning of the gesture within the \ac{SoI}.\\ 

\noindent\textbf{Data Gathering}
The collected dataset for the three distinct gestures (\textit{open hand}, \textit{two fingers}, and \textit{closed hand}) was conducted through a single volunteer. For each gesture, the hand remained static while the five configurations of the \ac{RIS} changed. This process was repeated 60 times (i.e., \textit{runs}), resulting in 900 samples (1 subject $\times$ 5 RIS configurations $\times$ 60 \textit{runs}). Additionally, noise measurements were taken without any hand in the \ac{SoI}. Specifically, ten \textit{runs} were performed for each of the five \ac{RIS} configurations, resulting in 50 noise measurements. Each sample is characterized by the number of frequency points $n_{res} = 201$, between \SI{5.0}{\giga\hertz} and \SI{6.5}{\giga\hertz}. 

During data capture, the signal varied slightly, which is expected given the difficulty in keeping the hand perfectly still. Considering that each \textit{run} included changing the configuration and saving the data, each \textit{run} took approximately 15 seconds, resulting in a total acquisition time of about 1 hour and 15 minutes (15 seconds $\times$ 5 \ac{RIS} configurations $\times$ 60 \textit{runs}), excluding re-setup and rest breaks. This data collection process was quite exhausting, making it impractical to use a real hand for collecting a large dataset required for \ac{HGR} approach.\\

\noindent\textbf{Experimental Results}
The obtained dataset was analyzed to observe if there were distinguishable differences between the hand gestures with this setup. Since the signal varied slightly due to the difficulty in maintaining a static hand position, averaging the signal would result in a loss of information. Therefore, the raw signals were analyzed. First, we examined the magnitude of the $S_{21}$ parameters for each gesture at a specific RIS configuration that steered the signal by 20 degrees. This approach allowed us to observe the variability between different \textit{runs} for the same gesture.

\begin{figure*}[t!]
\centering
\subfloat[phase response for no subject in space (i.e. noise)]{
\includegraphics{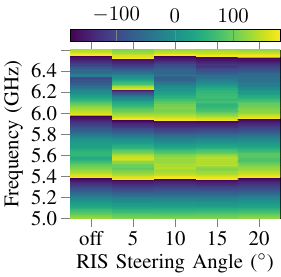}
\includegraphics[
trim={80 10 80 26}, clip, 
width=0.42\columnwidth]{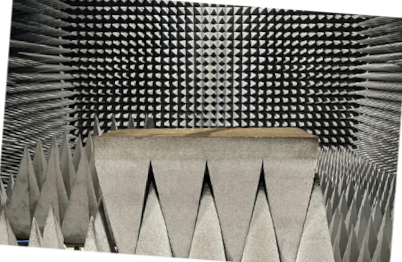}
\label{fig:pose1Phase}}
\subfloat[phase response for shown open hand gesture]{
\includegraphics{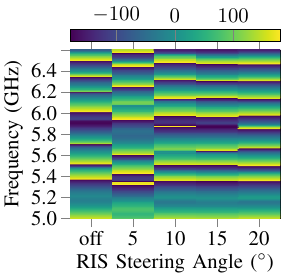}
\hfil
\includegraphics[
trim={80 40 80 8}, clip, 
width=0.42\columnwidth]{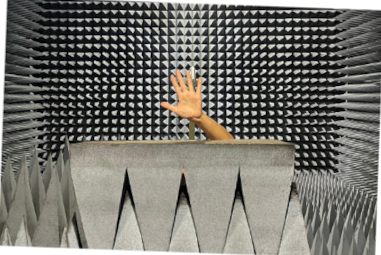}
\label{fig:pose2Phase}}

\subfloat[phase response for shown two finger gesture]{
\includegraphics{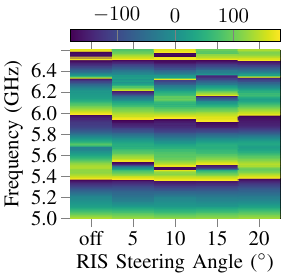}
\hfil
\includegraphics[
trim={80 30 80 25}, clip, 
width=0.42\columnwidth]{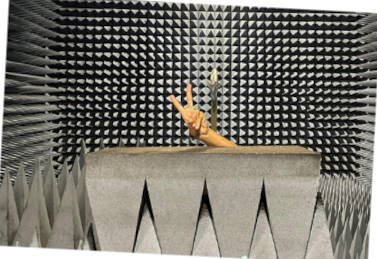}
\label{fig:pose3Phase}}
\subfloat[phase response for shown close hand gesture]{
\includegraphics{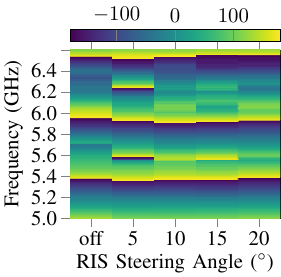}
\hfil
\includegraphics[
trim={90 46 120 45}, clip, 
width=0.42\columnwidth]{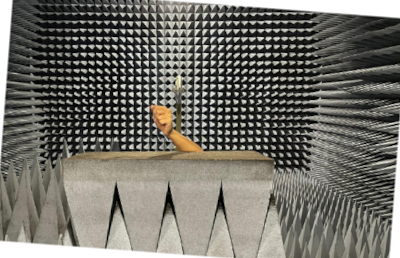}
\label{fig:pose4Phase}}
\caption{Phase responses for a frequency range of \SI{5.0}{\giga\hertz} to \SI{6.5}{\giga\hertz} for several \ac{RIS} configurations and hand gestures.}
\label{fig:medrealhand}
\end{figure*}

\Cref{fig:medrealhand} shows the phase response, derived from the respective $S_{21}$ parameters, for one measurement of each of the four hand gestures. The responses shown were randomly selected from a set of 60 measurements of the respective hand gesture (shown on the right-hand side). We show the phase response for five configurations of the \ac{RIS} for each gesture. Specifically, \emph{off}, i.e., purely specular, and for steering angles \SI{0}{\degree}, \SI{10}{\degree}, \SI{15}{\degree}, and \SI{20}{\degree}. For brevity, we omit the magnitude response but note that the same distinction in responses between gestures. We illustrate only one measurement per gesture since we observed that there was very little variation between measurements of the same gesture. From the data in \Cref{fig:medrealhand} it is evident that each gesture produces a distinct phase response, leading to the conclusion that there are clear differences in the received signal at the Rx for each gesture.\\

\noindent\textbf{Conclusion}
In summary, the $S_{21}$ parameters measured within the \ac{SoI} display distinct characteristics per hand gesture, which are consistent between measurements of the same gesture, and therefore the viability of the classification of gestures is demonstrated. Given the high effort of data gathering, we conduct the following experiment using a proxy hand model, instead of a human subject.

\subsection{RIS-based \acl{HGR} Classification}
\label{sub:hgrwood}

Having demonstrated that the $S_{21}$ parameters within the space of interest display variations with hand gestures, we gathered a larger dataset to allow for \ac{CNN} training and inference. The following sections present the data gathering setup, illustrate that the \ac{FCAO} based \ac{RIS} configuration provides more distinct $S_{21}$ responses relative to random phase control patterns, and show training and classification results for two different \ac{CNN} models.\\
%

\noindent\textbf{Experimental Setup}
Analogous to the previous experiment, \Cref{fig:setup_ris_wooden} depicts the setup with a wooden hand, where the Rx antenna is slightly tilted upwards (relative to \Cref{fig:setup_ris_real}) to focus on the center of the wooden hand. This adjustment ensures more similarity between both setups by isolating the region of interest, i.e., the hand and fingers, which was shown in the previous experiment to produce $S_{21}$ parameters capable of distinguishing gestures. That is, we wish to exclude potential response components from the wrist that compromise the differentiation in responses we observed in \Cref{sub:hgrhuman}.

Following our \ac{RIS} reconfiguration method (\Cref{sub:risconfig}) where configurations are grouped into \textit{frames}, we now consider measurement runs with 10 frames, where each frame contains 39 distinct \ac{RIS} configurations. These configurations were applied sequentially, resulting in a total of 390 measurements of the transmission coefficients $S_{21}$ per run. The experiment was conducted twice: once with random configurations and once with the optimized configurations determined by the \ac{FCAO} algorithm. The data  data acquisition system was adjusted to synchronize the configuration changes in the \ac{RIS} with the data collection. For each of the 39 configurations per frame, the \ac{VNA} was triggered to record the $S_{21}$ parameters, resulting in a consistent data capture of 390 total measurements per run.\\

\noindent\textbf{Data Gathering}
In this second experiment, a wooden hand was used for data collection. The collection process now applies the \ac{FCAO} algorithm, and therefore, the concept of frames will be discussed again. The copper tape was applied to the fingers to enhance reflection, as wood is a material with low reflectivity \cite{wood}. The same three hand gestures from the first experiment -- Open Hand, Two Fingers, and Close Hand -- were studied. For each gesture, 115 runs were conducted using random \ac{RIS} configurations and 115 runs with optimized configurations, both across 10 frames. Each sample contained transmission coefficients $S_{21}$ across the frequency range of \SI{5}{\giga\hertz} to \SI{6.5}{\giga\hertz}, with 201 points per sample, as in the first experiment. The time required to complete one run, which involved capturing data for all 390 configurations (39 configurations per frame $\times$ 10 frames), was about 3.5 minutes, for both the random and optimized configurations. The data acquisition time for the random configurations was 3.5 minutes$\times$ 115 runs = 6.7 hours per gesture, with the same duration for the optimized configurations. Considering the three different gestures, this resulted in a total acquisition time of around 39 hours. This extended acquisition time demonstrated the impracticality of using a real hand for large datasets, as it demands prolonged data collection, which is physically exhausting for participants.
The dataset created and used in this experimental campaign has been made publicly available\footnote{M. S. F. de B. Oliveira, F. M. Ribeiro, N. Paulino and L. M. Pessoa, “RIS Based Hand Gesture Recognition Dataset”. Zenodo, Sep. 12, 2024. doi: 10.5281/zenodo.13754235.}.\\

\begin{figure*}[t!]
\centering
{
    \captionsetup[subfigure]{
        format=hang, justification=raggedright,
        singlelinecheck=false, width=0.4\linewidth}
    \hspace{-8pt}
    \subfloat[magnitude and phase response for random configuration patterns and \textit{closed hand} gesture]{
        \includegraphics[
        height=0.205\linewidth]{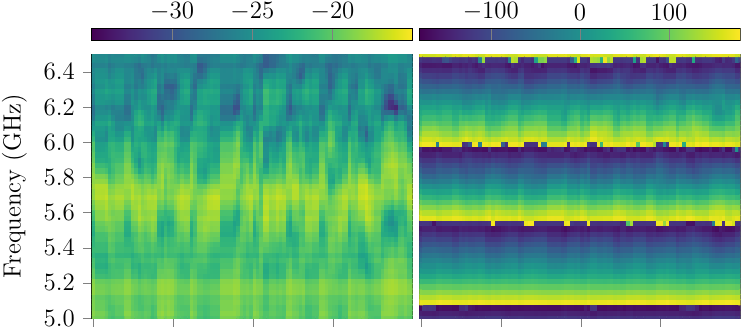}
        \label{fig:hgrwoodrandomclose}
    }
    \hspace{-9pt}
    \subfloat[magnitude and phase response for optimized configuration patterns and \textit{closed hand} gesture]{
        \includegraphics[
        height=0.205\linewidth]{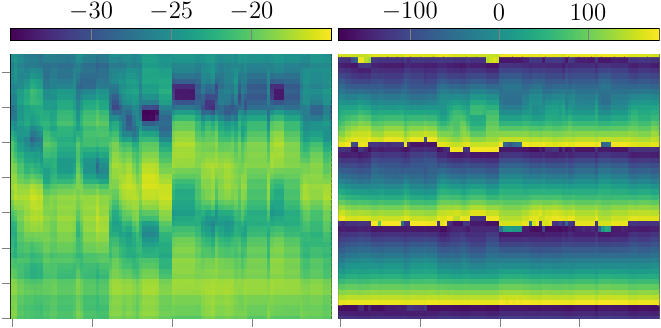}
        \label{fig:hgrwoodoptimclose}
    }
}
\hspace{-9pt}
\subfloat[]{\includegraphics[
trim={345 320 345 320}, clip,
width=0.10\linewidth]{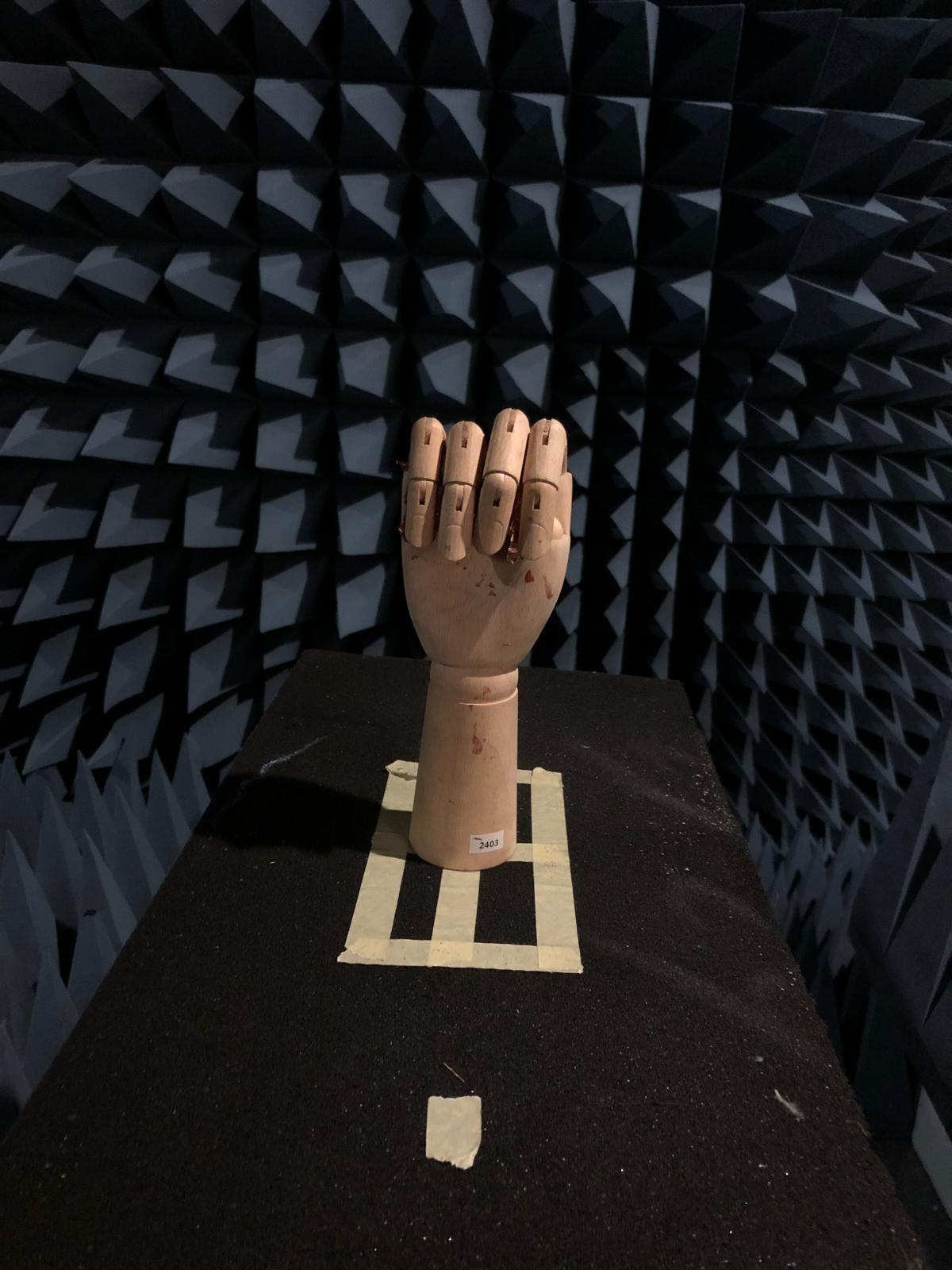}
\label{fig:hgrwoodclose}}
\vspace{5pt}

{
    \captionsetup[subfigure]{
        format=hang, justification=raggedright,
        singlelinecheck=false, width=0.4\linewidth}
    \hspace{-8pt}
    \subfloat[magnitude and phase response for random configuration patterns and \textit{open hand} gesture]{
        \includegraphics[
        height=0.172\linewidth]{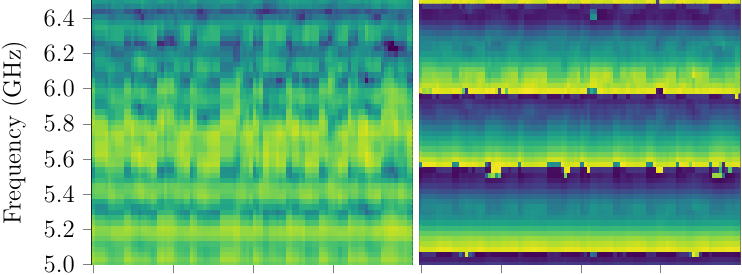}
        \label{fig:hgrwoodrandomopen}
    }
    \hspace{-9pt}
    \subfloat[magnitude and phase response for optimized configuration patterns and \textit{open hand} gesture]{
        \includegraphics[
        height=0.172\linewidth]{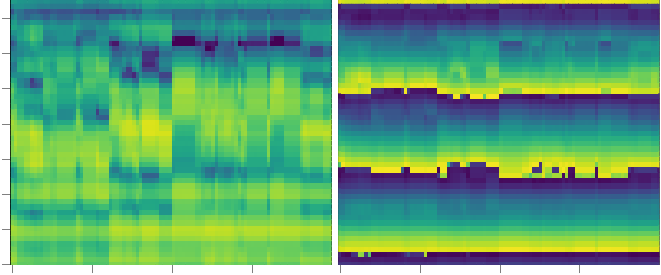}
        \label{fig:hgrwoodoptimopen}
    }
}
\hspace{-9pt}
\subfloat[]{\includegraphics[
trim={320 320 320 320}, clip, 
width=0.10\linewidth]{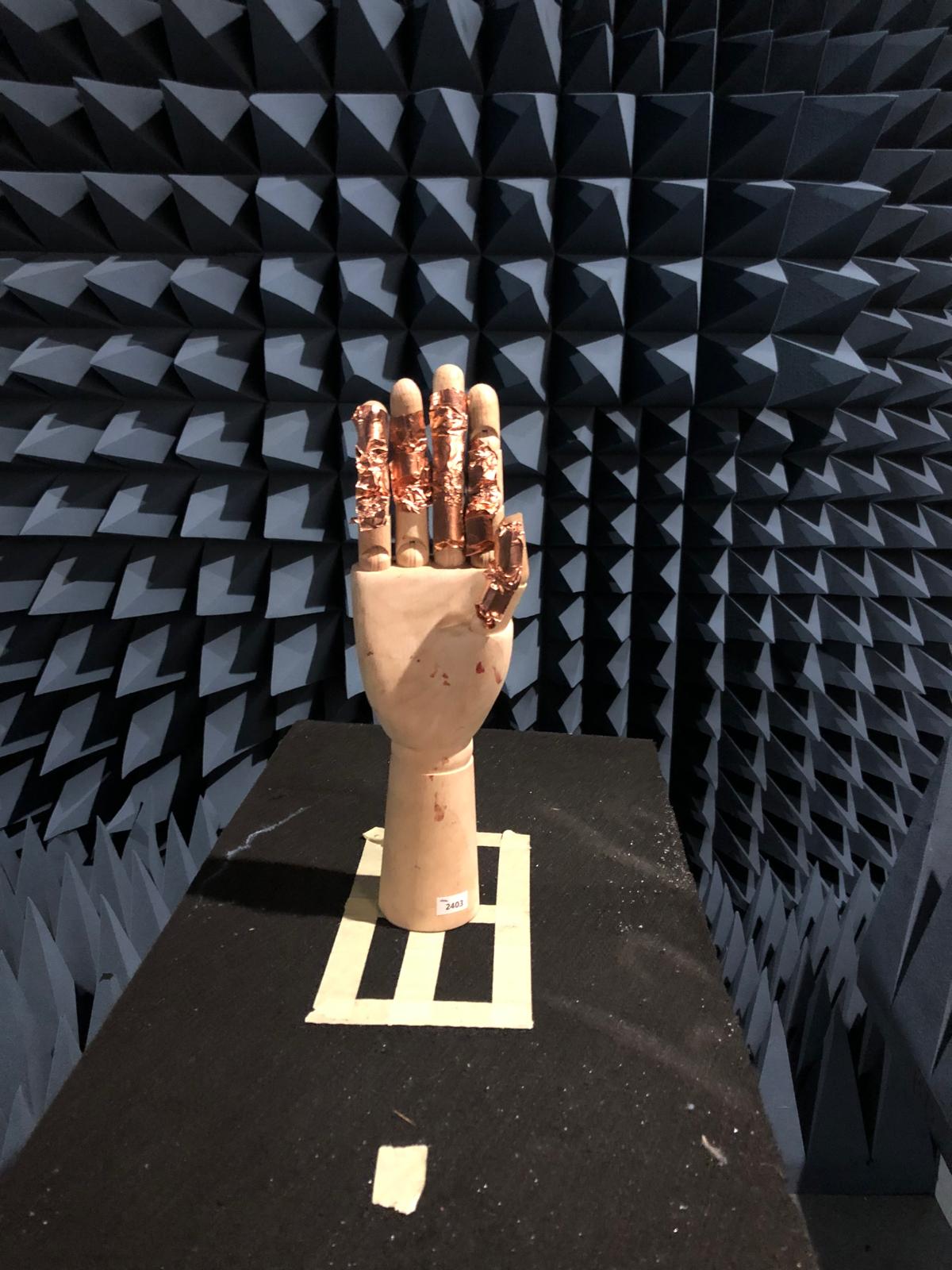}
\label{fig:hgrwoodopen}}
\vspace{5pt}

{
    \captionsetup[subfigure]{
        format=hang, justification=raggedright,
        singlelinecheck=false, width=0.4\linewidth }
    \hspace{-8pt}
    \subfloat[magnitude and phase response for random configuration patterns and \textit{two finger} gesture]{
        \includegraphics[
        height=0.21\linewidth]{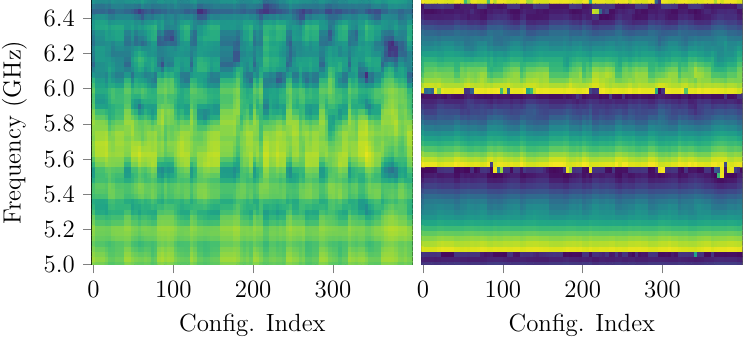}
        \label{fig:hgrwoodrandomtwo}
    }
    \hspace{-9pt}
    \subfloat[magnitude and phase response for optimized configuration patterns and \textit{two} gesture]{
        \includegraphics[
        height=0.21\linewidth]{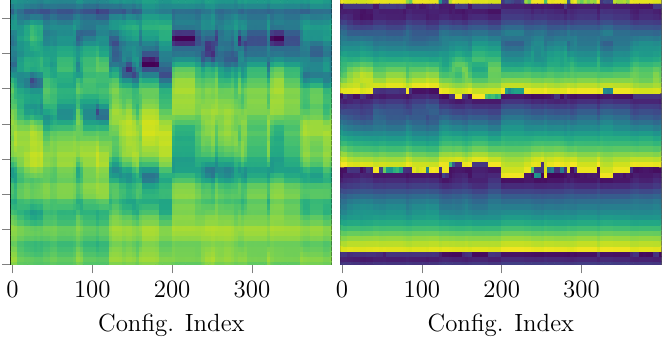}
        \label{fig:hgrwoodoptimtwo}
    }
}
\hspace{-9pt}
\subfloat[]{\includegraphics[
trim={380 340 380 340}, clip, 
width=0.10\linewidth]{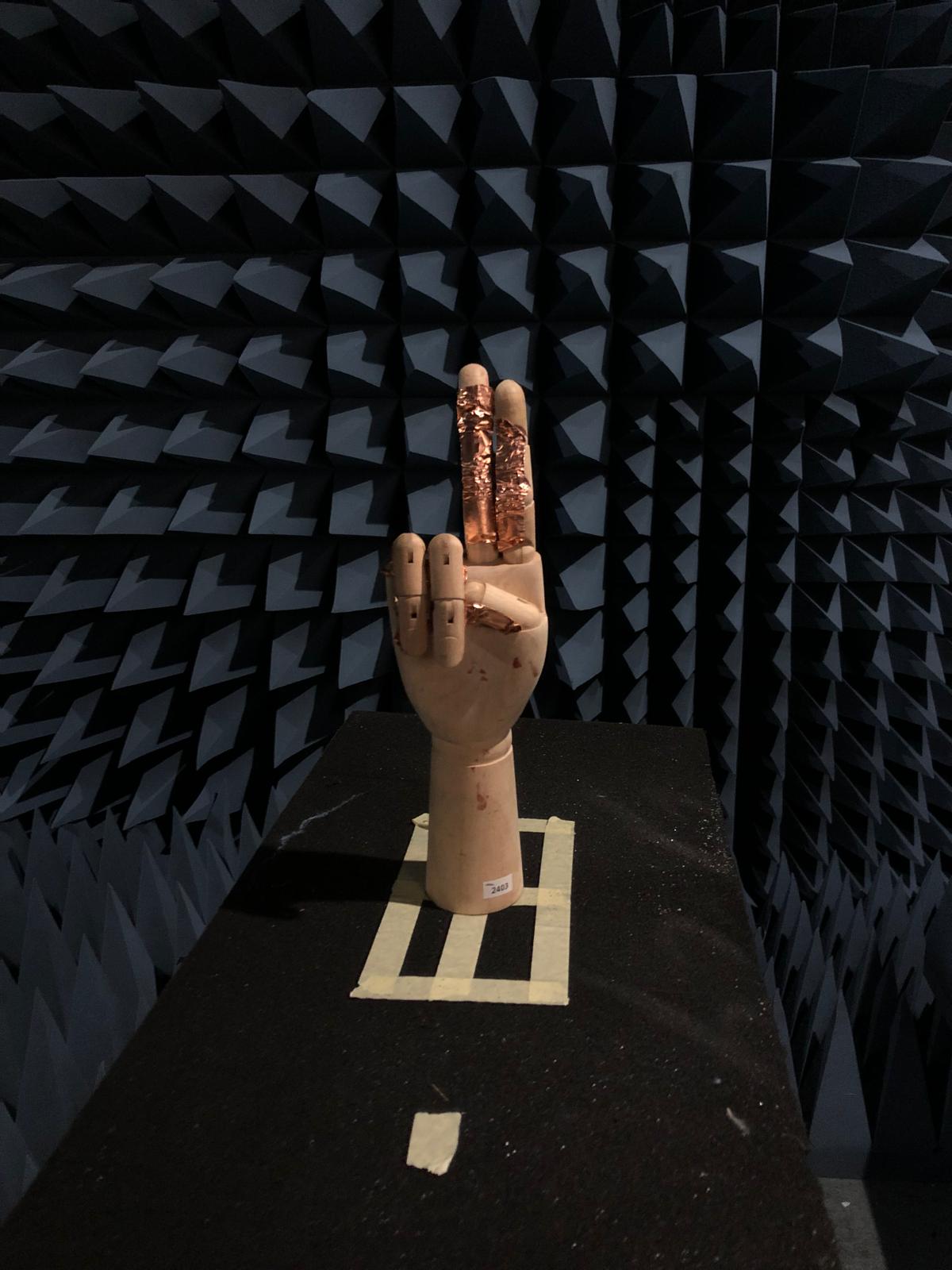}
\label{fig:hgrwoodtwo}}
\caption{Phase responses for a frequency range of \SI{5.0}{\giga\hertz} to \SI{6.5}{\giga\hertz} for random \ac{RIS} configuration sequences (\Cref{fig:hgrwoodrandomclose,fig:hgrwoodrandomopen,fig:hgrwoodrandomtwo}), and for the \ac{FCAO} optimized sequences (\Cref{fig:hgrwoodoptimclose,fig:hgrwoodoptimopen,fig:hgrwoodoptimtwo}), for each of the respective hand gestures (\Cref{fig:hgrwoodclose,fig:hgrwoodopen,fig:hgrwoodtwo})}
\label{fig:medwoodhand}
\end{figure*}

\noindent\textbf{Random vs. FCAO Optimized RIS Configurations}
We again plotted the collected $S_{21}$ data as images, mapping the responses in magnitude and phase for our frequency range, as a function of the sequence of \ac{RIS} configurations. \Cref{fig:medwoodhand} illustrates this for each gesture, for the random (left-hand side) and optimized \ac{RIS} (right-hand side) configurations. Axes ranges are identical for each subplot, and the horizontal axis represents the entire sequence of 390 configurations. Value ranges are shown by the color bars on top, with magnitude ranging between \SI{-35}{\decibel} to \SI{-15}{\decibel}, and phase between $\pm\SI{180}{\degree}$. The respective hand gestures for each row of plots are shown on the right.

For the random configuration patterns (\Cref{fig:hgrwoodclose,fig:hgrwoodopen,fig:hgrwoodtwo}), we observe little signal variability in magnitude across configurations, except for the last \textit{frame} of the configuration sequence, i.e., the very end of the sequence (configurations 351 to 390). The phase response also varies slightly throughout the configuration sequence. This suggests that having multiple frames with random configurations may not add much value, as the information across them is redundant. However, there is variability between gestures, particularly in the 
\SI{5.2}{\giga\hertz} to \SI{5.6}{\giga\hertz} range, where \textit{closed hand} results shows less variability relative to the \textit{open hand} gesture.

In contrast, the optimized patterns (\Cref{fig:hgrwoodoptimclose,fig:hgrwoodoptimopen,fig:hgrwoodoptimtwo}) reveal more variability in both magnitude and phase across frames, indicating that more information is captured from the space of interest when using the configuration pattern computed by our \ac{FCAO} implementation. Additionally, differences between gestures are more pronounced. For example, between \SI{6.0}{\giga\hertz} and \SI{6.5}{\giga\hertz}, the magnitude response is more attenuated relative to the random configuration sequence, and between the optimized cases, the attenuation is greater for \textit{closed hand} relative to \textit{open hand} in the same range.

For clarity, note that these plots show the average responses for each gesture, considering all 115 measurement runs per gesture. We present the averages directly, as opposed to analyzing specific chosen measurements runs per gesture since we observed very little variability for different measurements of the same gesture. For instance, considering the magnitude, and each set of 115 responses of the same set of measurements (i.e., same gesture and \ac{RIS} configuration sequence), we verify that the mean of the absolute differences of a gesture relative to the average response of that gesture ranges between \SI{0.2}{\decibel} and \SI{1.2}{\decibel}. Given the absolute range of \SI{-35}{\decibel} to \SI{-15}{\decibel}.


 
In other words, there is little variability in data between different runs for the same gesture, likely due to the anechoic environment and use of an immobile hand. Therefore, data augmentation techniques were employed to ensure a more robust dataset for training a neural network for hand gesture recognition.\\

\noindent\textbf{Data Augmentation for Machine Learning}
We augmented our data in two ways. Firstly, we had observed during the experiments on the human hand subject that slight adjustments or minor motion of the hand position was enough to cause response changes. Therefore, additional measurements were taken with the wooden hand placed in 8 different minor variations for each gesture. Given the conclusion regarding variability between measurement of the same gesture we have just presented above, we determined that repeating measurements per gesture did not significantly contribute to new information. Therefore, only a single measurement was taken for each new orientation. 
%
%

Instead, as our second technique for data augmentation, for each new variant position, we replicated the respective measurement 115 times, adding adding random white noise to each replica, with a mean of 0 and a standard deviation randomly chosen from [0.08, 0.1, 0.15, 0.2] as described in \cite{wn_radar}. This approach ensured a more robust dataset, allowing the neural network to learn patterns beyond the limited original data and avoid overfitting. This process resulted in a total of 1035 samples per gesture for both random and optimized configurations (115 runs $\times$ 9 orientations). With three gestures, the final dataset comprised 3105 samples for random configurations and 3105 samples for optimized configurations, where each sample corresponds to an image used for \ac{CNN} learning and classification.\\

\noindent\textbf{CNN Classification with Model \#1}
The first \ac{CNN} model was adapted from the one presented by Hu \textit{et al.} \cite{9133157}, from where we also based our \ac{FCAO} implementation. We employ the same model here to arrive at a comparison of performance given the different input data collected. Specifically, in this reference paper, only the information at the operating frequency of the \ac{RIS} was used for classification. In our case, due to the characterization results of the unit cell and \ac{RIS}, this frequency $f_c = \SI{5.91}{\giga\hertz}$ as we explained above (\Cref{sub:unitcellcharacterization}. Thus this model relies only on a single frequency and furthermore takes only the average of the response per each configuration frame (i.e., it considers the average of the responses of all configurations within a frame). In our case, and considering the inputs are the only real components of the complex values of the $S_{21}$ parameters, resulting in in a feature vector with dimension $(1, 10)$. 

Regarding architecture, the neural network consists of three fully connected layers: an initial layer with 64 neurons, a hidden layer with 32 neurons, both using the ReLU activation function and an output layer with \textit{softmax} activation. 
The dataset was split into \SI{80}{\percent} for training, \SI{10}{\percent} for validation, and \SI{10}{\percent} for testing. We implemented the same model and trained it with both of our data subsets (i.e., $S_{21}$ data for random and optimized configuration sequences).


\begin{figure}
\includegraphics{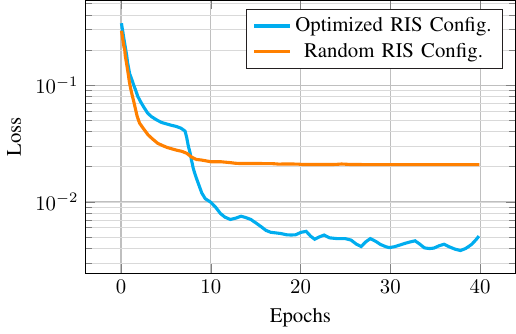}
\caption{Training epochs for CNN model \#1, for both types of collected data} 
\label{fig:converge_c1}
\end{figure}

During training, shown in \Cref{fig:converge_c1}, it is evident that in the initial epochs, the model converges faster with the random configurations than with the optimized ones. This suggests that the optimized configuration contains more complex information, making it harder to learn at first. However, as training progressed the optimized configuration showed improved accuracy, achieving around $5\times$ smaller loss. This pattern was also observed in the study by Hu \textit{et al.} \cite{9133157}. 

\begin{figure}[t!]
\centering
{
\subfloat[CNN \#1 and random config.]{
\includegraphics{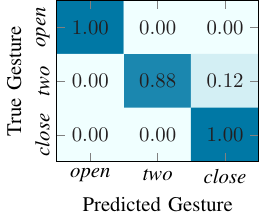}
\label{fig:confuse1}}
\hfill
\subfloat[CNN \#1 and optimized config.]{
\includegraphics{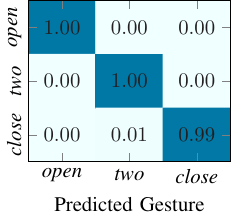}
\label{fig:confuse2}}
}
\caption{Gesture recognition accuracy for the \ac{CNN} architecture utilizing only $S_{21}$ information for a frequency of \SI{5.91}{\giga\hertz} using both random configurations or the \ac{FCAO} optimized configurations}.
\label{fig:class_res_cnnSimple}
\end{figure}

\Cref{fig:class_res_cnnSimple} presents the confusion matrices for both configurations, demonstrating an advantage for the optimized setup. This confirms that the optimized configuration matrix, designed with low average mutual coherence, can achieve superior accuracy in gesture recognition. Specifically, the optimized configuration resulted in a \SI{99.67}{\percent} recognition accuracy, providing a \SI{3.53}{\percent} improvement compared to the random configuration, which achieved \SI{96.14}{\percent}.\\

\begin{figure*}[t!]
  \centering
  \includegraphics[width=0.85\linewidth]{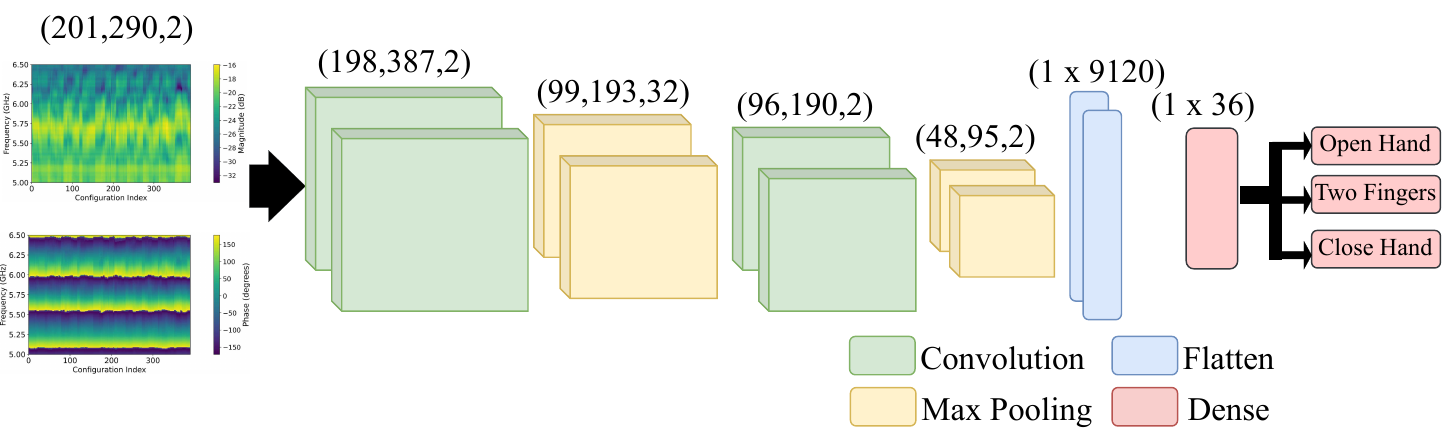}
  \caption{Architecture of the \ac{CNN} developed for gesture classification (Open Hand, Two Fingers, Close Hand). The input to the network consists of the magnitude and phase of the S$_{21}$ parameters.}
  \label{fig:arquitetura2}
\end{figure*}

\noindent\textbf{CNN Classification with Model \#2}
The architecture of our second \ac{CNN} is presented in \Cref{fig:arquitetura2}. We now consider two channels, the magnitude and phase responses, represented as the image plots we have previously shown. That is, this model considers broadband information from \SI{5.0}{\giga\hertz} to \SI{6.5}{\giga\hertz}, and also the response of each individual configuration of the sequence, rather than an average response per frame. Thus the resulting dimension of the input layer is two feature vectors with dimensions $(201,390)$, and the output layer consists of three nodes for the hand gestures to be classified. This architecture allows the network to learn features from both the magnitude and phase components of the $S_{21}$ parameters, potentially capturing distinct and complementary information.

\begin{figure}
\includegraphics{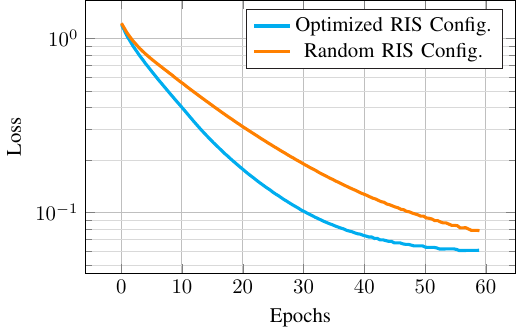}
\caption{Training epochs for CNN model \#2, for both types of collected data} 
\label{fig:converge_c2}
\end{figure}

\begin{figure}[t!]
\centering
{
\subfloat[CNN \#2 and random config.]{
\includegraphics{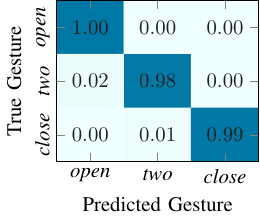}
\label{fig:confuse3}}
\hfill
\subfloat[CNN \#2 and optimized config.]{
\includegraphics{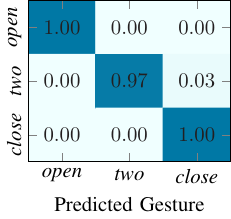}
\label{fig:confuse4}}
}
\caption{Gesture recognition accuracy for the \ac{CNN} architecture utilizing broadband information from \SI{5.0}{\giga\hertz} to \SI{6.5}{\giga\hertz} using both random configurations or the \ac{FCAO} optimized configurations}.
\label{fig:class_res_cnnComplex}
\end{figure}

The training for this network is shown in \Cref{fig:converge_c2}, once again for both types of configuration strategies. In this case, the optimized case produces a lower loss per epoch even from the start of the training process. However, both configuration methods converge to a similar loss. That is, even though the gains from \ac{FCAO} are less evident for this model and respective input data, its use still permits a faster learning convergence.

The resulting classification accuracies for both of our data subsets (i.e., random and optimized sequences) are shown in \Cref{fig:class_res_cnnComplex}. Both the random and optimized configurations achieved high recognition accuracies. The average accuracy for the dataset retrieved using random configurations is \SI{99.047}{\percent}, which increases minimally to \SI{99.057}{\percent} for the optimized configuration dataset. This smaller increase relative to the previous \ac{CNN} model again suggests that optimizing the \ac{RIS} configuration yields little benefit, if also combined with data retrieved over a broader range of frequencies. 


We can briefly justify this based on how the formulation of the channel and the \ac{FCAO} algorithm operate. Underlying the approach are compressive sensing principles, where the \ac{FCAO} algorithm can provide benefits if the number of measurements $K$ is ideally much smaller than the number of distinct spatial regions of interest $M$. In other words, for the first \ac{CNN} model with an input size of $(1, K=10)$ for our case where $M=32$, these conditions are minimally satisfied. 

However, in the second approach, by not averaging the signals across frames, we leverage all 390 distinct signals, which results in $K=390$, greatly exceeding $M$. This substantial increase in measurements makes the application of the \ac{FCAO} algorithm less impactful regarding the increase of information extracted from the channel. Both configurations achieve similarly high performance, demonstrating that with such a large number of measurements, the choice of configuration has minimal impact on the classification results. 

In other words, while the use of an optimized configuration using \ac{FCAO} increases the information captured from a narrowband signal, at the cost of the computational effort of determining this configuration, we achieve a similar level of information by capturing magnitude and phase data for a broader frequency band.\\ 

\noindent\textbf{Conclusion}
We have successfully implemented an \ac{RF}-based \ac{HGR} approach, relying on our own \ac{RIS} design. We have shown that $S_{21}$ data for a bandwidth of \SI{1.5}{\giga\hertz} within the WiFi-6E range provides sufficient information for gesture classification, even for a \ac{RIS} design with only two control states per unit cell. 

Regarding maximizing the information that can be extracted from the channel, Hu \emph{et al.} \cite{9133157} demonstrated this through the \ac{FCAO} algorithm for a 2-bit \ac{RIS}, achieving a \SI{14.6}{\percent} accuracy improvement in their \acf{HPR} setup, relative to random configurations sequences. Although not directly comparable, we achieve an analogous \SI{3.53}{\percent} increase in classification accuracy for our 3 gestures in our \ac{HGR} setup using the same approach. The difference in accuracy increase can likely be attributed not just to different setups, but especially to the increased number of possible configuration states in a 2-bit \ac{RIS}, and respective control of the \ac{RF} medium. However, we have demonstrated that optimized configurations can still improve performance even for a 1-bit \ac{RIS}, a smaller \ac{SoI}, and different classification task. Moreover, we have shown that equivalent classification performances can be achieved even with a random sequence of \ac{RIS} configuration states if a larger amount of \ac{RF} data is directly collected from the channel.

\section{Conclusion}
\label{sec:conclusion}

In this work, we have demonstrated a \acf{HGR} approach that is capable of classifying three distinct human hand gestures, relying only on \ac{RF} data retrieved from a channel including the \ac{SoI} within which the human hand target is located. The channel is a so called \textit{smart radio} environment, as it includes a \acf{RIS} with which the response of the channel can be controlled. In this way, the information extracted from the \ac{SoI} enriched, since different \ac{RIS} configurations result in distinct $S_{21}$ parameters. 

We collected and published a large data set for \ac{HGR} in an anechoic chamber setup, for three possible human hand gestures, and for a frequency range of \SI{5.0}{\giga\hertz} to \SI{6.5}{\giga\hertz}. For two \acf{CNN} models, and two strategies for \ac{RIS} reconfiguration, we have demonstrated a classification accuracy above \SI{90}{\percent}. Furthermore, we have demonstrated this using our own hardware designs for a \ac{RIS} tile, and supporting control eletronics, targeting the WiFi-6E range. Thus, we have advanced the state of the art on \ac{RF} based solutions for \acf{HAR}, specifically within the context of emergent sixth generation communications.

Our \ac{RIS} design is a 8x8 tile using a PIN diode based unit cell was validated in anechoic chamber environment, where we analyzed the performance for several frequencies and steering angles. Despite a deviation from the intended central frequency, we demonstrated successful steering capabilities at \SI{6.16}{\giga\hertz}, for a signal emitted by a feed horn located \SI{40}{\centi\meter} away from the \ac{RIS}. Our design will be available as open-source, accelerating research on \ac{RIS} implementations and applications for WiFi-6E networks.

Regarding future work from a sensing perspective, we are assembling a setup with four tiles in order to achieve a total of 256 units cells. With this larger \ac{RIS}, we expect that the \ac{HGR} approach can be improved by potentially quadrupling the information from the channel. Due to a larger \ac{RIS}, we can also consider a larger \acf{SoI}, and thus perform full body \ac{HPR}. From a hardware design and communication perspective, We will also evaluate the improved steering capabilities with this larger \ac{RIS}, and will also explore the multi-beam capabilities afforded by a tile level control. Finally, we will iterate on the unit cell design to mitigate deviations from the intended central frequency, which may include focusing on a 2-bit control, and also consider \ac{RIS} designs for FR2 frequencies. 


\balance

\bibliographystyle{IEEEtran}
\bibliography{refs}
\end{document}